\documentclass[a4paper,11pt]{article}
\pdfoutput=1 

\usepackage{jheppub} 

\usepackage[T1]{fontenc} 

\usepackage{graphicx}
\usepackage{grffile}  
\usepackage{xspace}      

\usepackage{xcolor}

\title{\boldmath Optimising top-quark threshold scan at CLIC using genetic algorithm}

\author{K.~Nowak}
\author{A.~F.~\.Zarnecki}
\affiliation{Faculty of Physics, University of Warsaw, Pasteura 5, 02-093 Warszawa, Poland}

\emailAdd{k.nowak27@student.uw.edu.pl}
\emailAdd{zarnecki@fuw.edu.pl}

\newcommand{\epem}{\ensuremath{\text{e}^+\text{e}^-}\xspace}
\newcommand{\fbinv}{\ensuremath{\text{fb}^{-1}}\xspace}
\newcommand{\abinv}{\ensuremath{\text{ab}^{-1}}\xspace}
\newcommand{\ttbar}{\ensuremath{\text{t}\bar{\text{t}}}\xspace}
\newcommand{\qqthr}{\texttt{QQbar\_threshold}\xspace}
 
\newcommand{\textdef}[1]{\emph{#1}} 

\abstract{
One of the important goals at the future \epem colliders is to measure
the top-quark mass and width in a scan of the pair production threshold.
However, the shape of the pair-production cross section at the
threshold depends also on other model parameters, as the top Yukawa
coupling, and the measurement is a subject to many systematic uncertainties.
Presented in this work is the study of the top-quark mass
determination from the threshold scan at CLIC.
The most general approach is used with all relevant model parameters
and selected systematic uncertainties included in the fit procedure.
Expected constraints from other measurements are also taken into
account.
It is demonstrated that the top-quark mass can be extracted with
precision of the order of 30 to 40\,MeV, including considered
systematic uncertainties, already for 100\,\fbinv of data collected at
the threshold.
Additional improvement is possible, if the running scenario is
optimised.
With the optimisation procedure based on the genetic algorithm  the
statistical uncertainty of the mass measurement can be reduced by
about 20\%.
Influence of the collider luminosity spectra on the expected
precision of the measurement is also studied.
}


\graphicspath{ {./plots/} }

\begin{document} 

\maketitle
\flushbottom


\section{Introduction}

The Compact Linear Collider (CLIC) is a linear \epem collider
project, that is considered as a possible next large infrastructure at
CERN~\cite{clic-cdr,clic-pip}.  
CLIC is proposed as a staged machine running at energies from 380\,GeV
up to 3\,TeV. 
The physics program of the first CLIC energy stage, at 380\,GeV,
focuses on precise measurements of the Higgs boson properties
including its couplings to other SM particles \cite{clic-higgs} and a
study of top-quark production and decays, incorporating an energy scan
over the \ttbar production threshold \cite{clic-top}.

Precise measurement of the top-quark mass is essential  for the
understanding of the Higgs mechanism, electroweak symmetry breaking
and for constraining many ``new physics'' scenarios. 
Scanning the threshold for top-quark pair production,
$\epem \to \ttbar$, 
was proposed as the method for top-quark mass measurement
even before the top quark was actually
discovered~\cite{Fadin:1987wz,Fadin:1988fn,Strassler:1990nw}.    
As the production cross-section can be calculated with
a high degree of precision using theoretically well-defined
top mass definitions~\cite{Hoang:1999zc}
it is currently assumed to be the most precise method for top-quark
mass determination and least sensitive to theoretical uncertainties.

\begin{figure}[b]
\centering
\includegraphics[width = 0.5\textwidth]{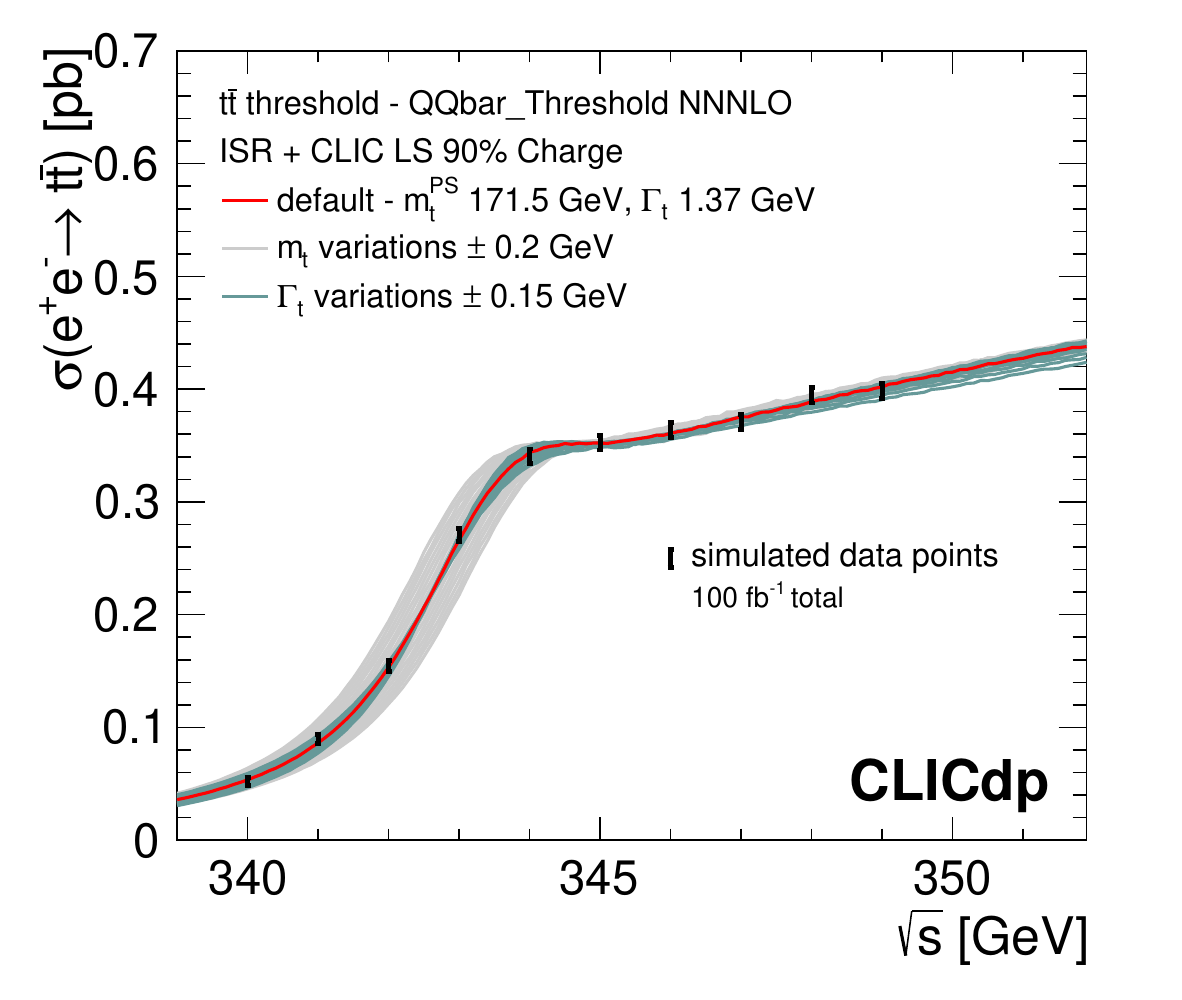}
\caption{Illustration of a top-quark threshold scan at CLIC with a
  total integrated luminosity of 100\,\fbinv.
The bands around the central cross section curve show the dependence
of the cross section on the top-quark mass and width, the error bars on the
simulated data points show the expected statistical uncertainties of
the cross section measurement. Figure taken from \cite{clic-top}.}
\label{fig:base}
\end{figure}

Scan scenario with ten  energy points separated by 1\,GeV, with
10\,\fbinv of data collected at each energy is considered as a
baseline scenario for CLIC, as shown in Fig.~\ref{fig:base}.
Detailed study \cite{clic-top} showed that expected statistical
uncertainty on the mass is around 21\,MeV and on the top-quark width -
51\,MeV.
However, systematic uncertainties are expected to limit the ultimate
precision.
In particular, mass uncertainty originating from the uncertainty of
the strong coupling is estimated to be about 30\,MeV
\cite{Simon:2016htt}.
The combined theoretical, parametric and experimental systematic
uncertainties are expected to be in the range 40\,MeV to 70\,MeV,
depending on the assumptions \cite{clic-top}.

While these results are very encouraging, they were obtained from the
threshold scan fit taking only the two model parameters, top-quark
mass and width, into account. 
Yet, threshold cross-section shape depends also on other parameters,
as the top Yukawa coupling and the strong coupling constant.
Fit results are also sensitive to the normalisation of the model
predictions and the background level estimates. 
The main goal of the presented study was to quantify the influence of
additional model parameters, and related uncertainties, on the
precision of top-quark mass determination at CLIC.
By including corresponding parameter variations in the fit, influence
of the considered systematic uncertainties can also be reduced.

As the baseline running scenario seems to be conservative, the
additional objective of the study was to investigate to what extent
statistical uncertainties can be reduced when using the optimised
running scenario.
The scan optimisation is possible only if the top-quark
mass is already known to $\mathcal{O}$(100\,MeV).
If this level of precision is not reached by earlier measurements,
an initial scan with fewer energy points can be required, with only a
small fraction of total integrated luminosity dedicated to the
threshold scan, see eg.~\cite{Simon:2019axh}.
However, suggesting any realistic running scenario, when the scan
sequence is adjusted basing on the already collected scan data, is
a much more complex problem and is beyond the scope of this work.
Results included in this paper supersede results presented previously
in \cite{wilga2018,wilga2019,thesis,ichep2020}.


\section{Modelling of the threshold scan}

In this chapter the procedure developed to model the
top-quark pair production cross section measurements in a threshold
scan at CLIC is described.
Two elements are used to calculate the expected cross section values:
theoretical predictions given in terms of the cross section templates
and the expected luminosity spectra for CLIC.  
    
\subsection{Cross-section templates}
    
\begin{figure}[tb]
\includegraphics[width=0.49\textwidth]{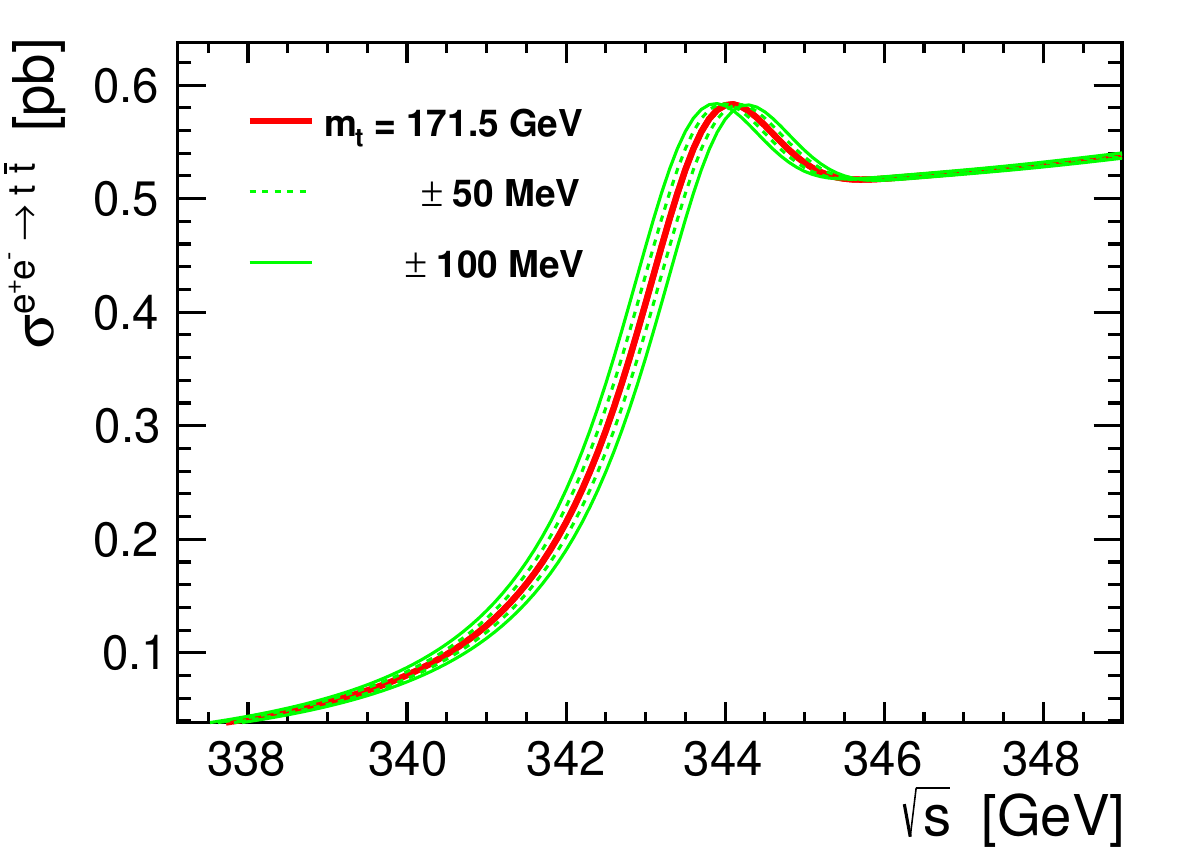}
\includegraphics[width=0.49\textwidth]{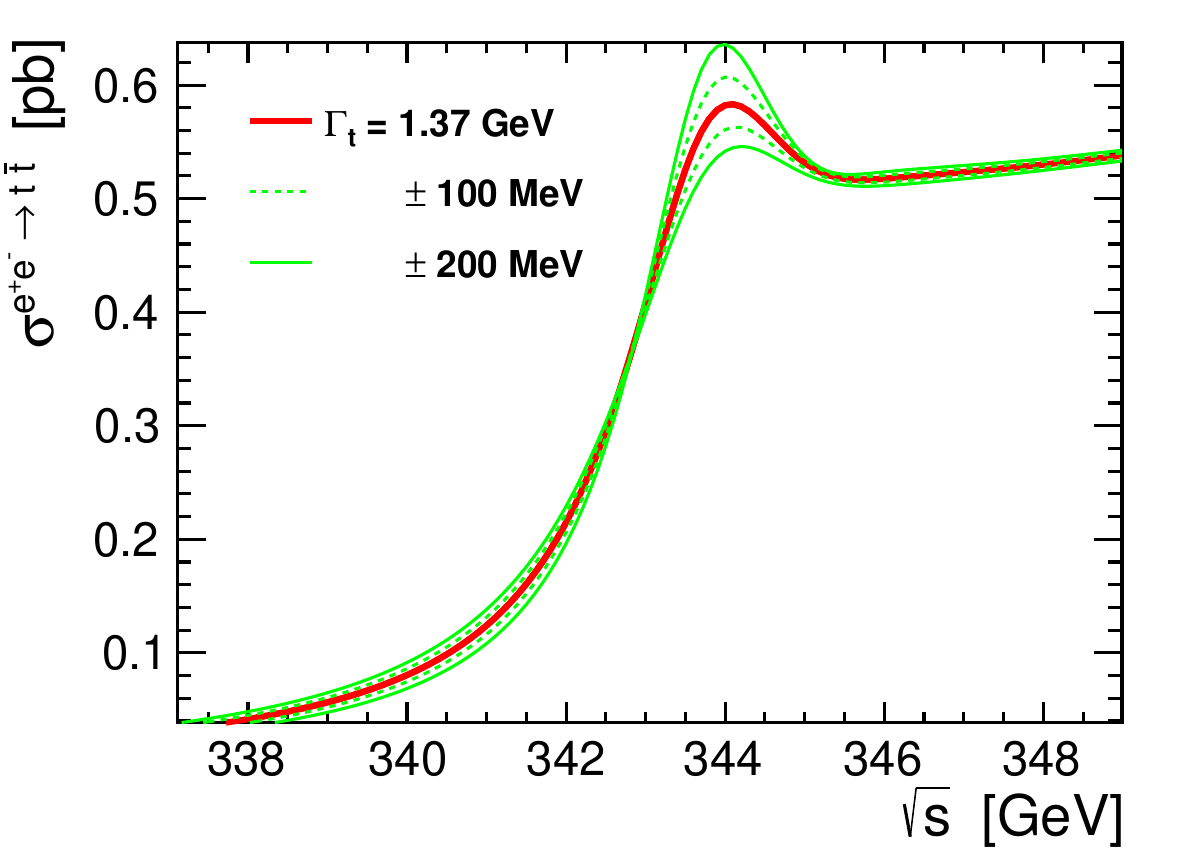}
\caption{Examples of cross-section templates generated using
  \qqthr software for different values of top-quark mass
  (left) and width (right) \cite{kacper_debski}.} 
\label{fig:templates}
\end{figure}

Considered in the presented study are the cross section templates
generated, assuming different values of
top quark parameters, using 
\qqthr  software \cite{qq,qq2}.
Parameters used as an input to \qqthr calculations are listed in the
upper part of Tab.~\ref{tab:qq_par}. 
\begin{table}[tb]
\centering
\begin{tabular}{||l c|c|c|c|c||}
\hline
\multicolumn{2}{||c|}{Parameter} & Nominal & Min. & Max. & Step \\
\hline
top-quark mass &  $m_t$ &  171.5\,GeV & 171.4\,GeV &  171.6\,GeV & 25\,MeV \\
top-quark width & $\Gamma_t$ & 1.37\,GeV & 1.17\,GeV & 1.57\,GeV & 0.05\,GeV \\
top Yukawa coupling & $y_t$ & 1.0 &  0.6 & 1.4 & 0.1 \\
strong coupling constant & $\alpha_s$ & 0.1185 & 0.1145 & 0.1225 & 0.001 \\
overall renormalisation scale & $\mu$ &  80\,GeV &    &   &   \\
nonresonant contr. scale & $\mu_w$ &  350\,GeV &  &  &  \\
\hline
data normalisation & $\alpha$  & 1  &  &  &   \\
background contribution & $f_{bg}$ & 73\,fb & 61\,fb & 85\,fb & 3\,fb  \\
\hline
\end{tabular} 
\caption{Parameters used as an input to \qqthr (upper part of the
  table) and normalisation uncertainties considered in the analysis
  (lower part).
Top-quark Yukawa coupling, $y_t$, is given relative to SM
predictions. Overall renormalisation scale, $\mu$, and the energy
scale for nonresonant contributions, $\mu_w$, were not varied in the
described study. Range and step values are also not indicated for the
overall data normalisation factor $\alpha$, as it is evaluated
analytically in the fit procedure (see Section \ref{sec:fit_method}
for details).
} 
\label{tab:qq_par}
\end{table}
Nominal cross section template was first generated for nominal values
of all parameters. 
Three top-quark parameters considered in the study:  mass, $m_t$,
width, $\Gamma_t$,
and Yukawa coupling relative to SM prediction, $y_t$,
as well as the strong coupling constant, $\alpha_s$, 
were then varied in small steps, 
corresponding to the expected experimental sensitivities, from their
nominal values, resulting in additional cross section templates.
Total of 274 templates were generated for different variations of
model parameters.
Each of those templates consists of 300 points representing
cross-section for energy in range from 330 to 360\,GeV.  
Examples of cross-section templates are shown in Fig.~\ref{fig:templates}.
Initial state radiation was included in the template cross section
calculations.
                
\subsection{Luminosity spectra}
                
In order to take into account environment of the experiment, theoretical
cross section templates, as generated with the \qqthr, were convoluted
with the expected CLIC luminosity spectra to obtain final cross-section
templates used in the analysis.  
\begin{figure}[tb]
      \includegraphics[width=0.50\textwidth]{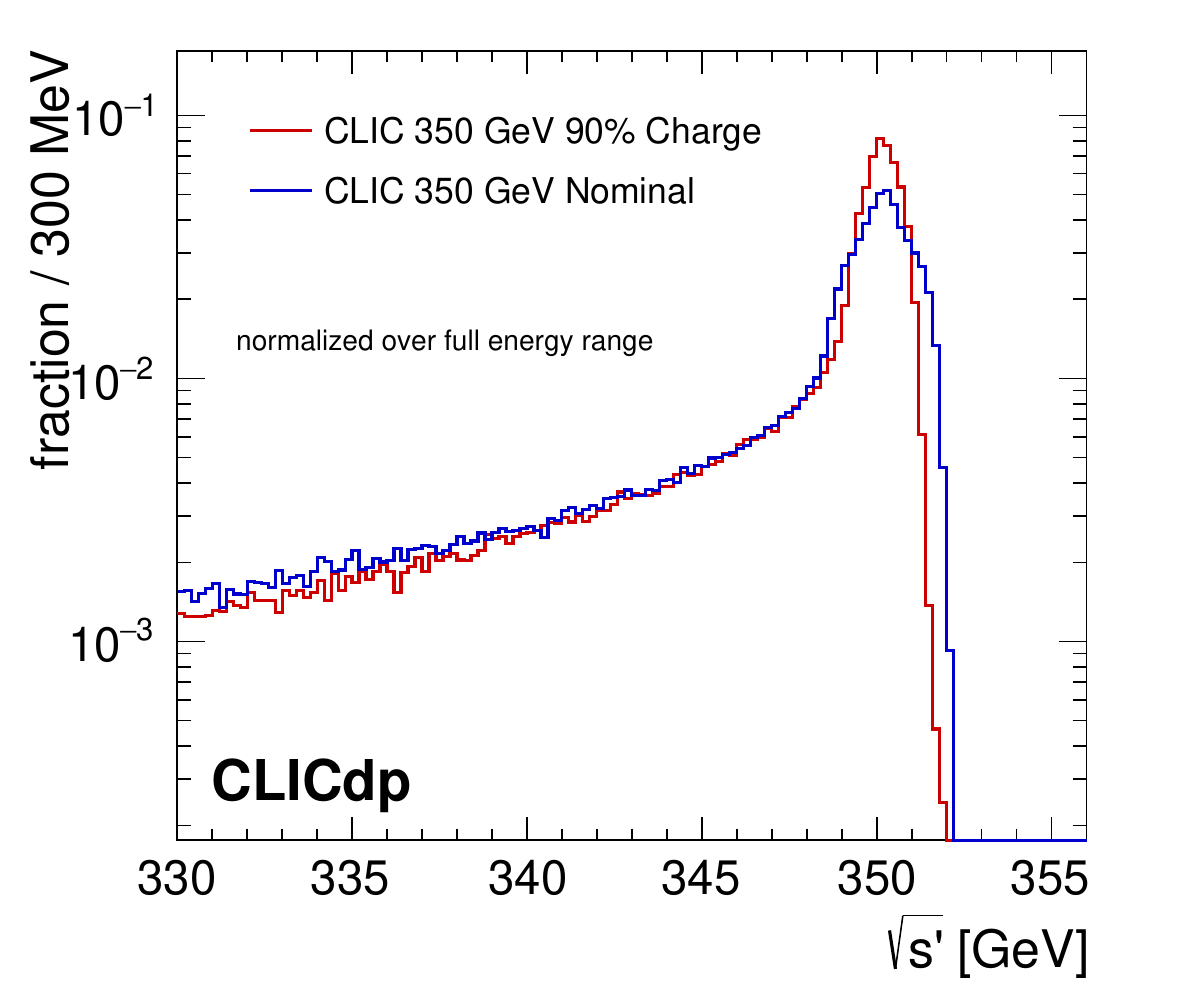}
      \includegraphics[width=0.48\textwidth]{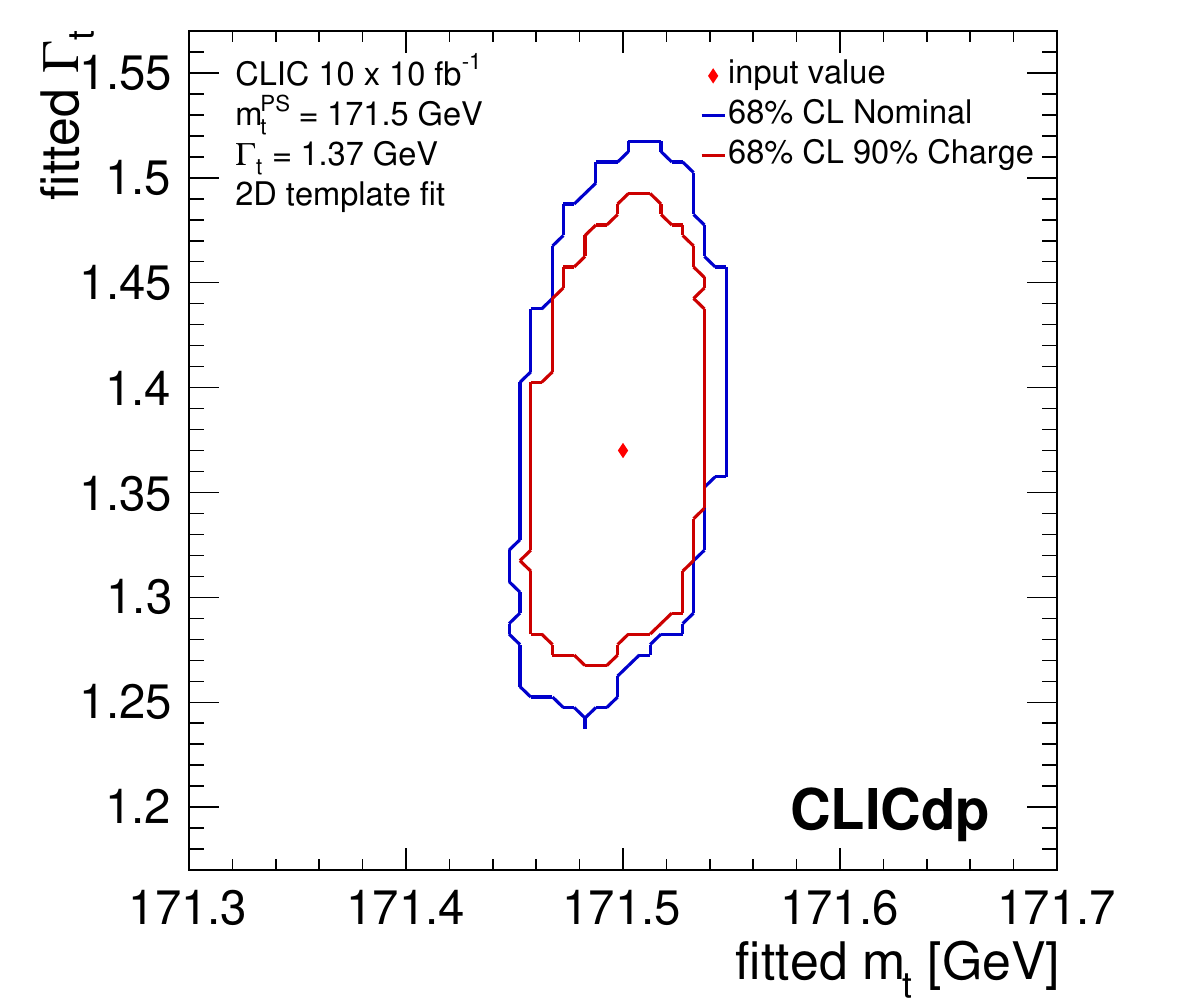}
 \caption{Left: two scenarios of the CLIC luminosity spectrum for a
threshold scan, one based on the nominal accelerator parameters
(Nominal) and one optimised for reduced beamstrahlung (90\% charge).
Right: 68\% CL statistical uncertainty contours of  the top-quark mass
and width fits to the top threshold scan data, for two considered
liminosity spectra scenarios, assuming an integrated luminosity of
100\,\fbinv in both cases. Figures taken from \cite{clic-top}.} 
                    \label{fig:lum}
\end{figure}
Two luminosity spectra were considered for the threshold scan: the
nominal spectra expected for the first stage of CLIC, and the one
with reduced bunch charge (90\% charge), which allows to obtain
narrower energy spectra at the cost of reduced instantaneous
luminosity.
Previous study \cite{clic-top} showed that the reduced
charge option results in smaller statistical uncertainties of the
threshold fit, see Fig.~\ref{fig:lum}, and this spectra was therefore
used for the presented study.

\subsection{Reference scenario}

At the first energy stage CLIC is assumed to run at the energy of 380\,GeV, which 
was selected for optimising both the Higgs boson and top quark measurements.
With 1\,\abinv of total integrated luminosity at this stage, a
dedicated scan of 100\,\fbinv is planned at the \ttbar
threshold.
As already mentioned above, the baseline scenario of the threshold
scan assumes running at 10 equidistant energy points taking 10\,\fbinv
of data for each value of $\sqrt{s}$.
When generating simulated measurements (pseudo-data), the overall
top-pair event reconstruction efficiency of 70.2\% (including
the branching fractions of the considered fully-hadronic and
semi-leptonic top pair decays) was assumed, and the background
contribution remaining after the event selection procedure
corresponding to the cross section of 73\,fb~\cite{Seidel}.
Reconstruction efficiency is assumed to be independent of the
collision energy (in the considered energy range) and of the model
parameters. 

\subsection{Systematic uncertainties}

In addition to the variations of the \qqthr input parameters,
as listed in the upper part of Tab.~\ref{tab:qq_par},
possible variations of the overall data normalisation, $\alpha$,
and of the assumed background contribution, $f_{bg}$, were also
considered as additional parameters in the modelling of the
threshold scan results. 
Data normalisation factor $\alpha$ covers possible systematic
variations due to the luminosity measurement, event selection
efficiency or normalisation of the theoretical predictions.
Uncertainties on the overall data
normalisation, $\Delta$, on the background normalisation,
$\delta_{bg}$, and on the value of the strong coupling
constant, $\sigma_{\alpha}$ are considered as the sources of the
systematic uncertainties in the extraction of the top-quark
mass and other model parameters,
and corresponding constraints were included in the fit to the
  threshold scan data. 
If not stated otherwise, relative uncertainty of 1\% is assumed for
the overall data normalisation~\cite{clic-cdr}, 
while uncertainty of 2\% is used for the background contribution.
Strong coupling constant, $\alpha_s$, is assumed to be known to 0.001.

Additional systematic uncertainties in the modelling of the threshold
cross section can be due to variation of the renormalisation scale, $\mu$, 
assumed in the \qqthr calculations and to the uncertainty of the
nominal top-quark mass, $m_t$, assumed when planning the threshold scan.
These uncertainties were not taken into account in the fit procedure
and their influence on the fit results is discussed in
Sec.~\ref{sec:results}.   

Top-quark mass parameter of the \qqthr program is defined in the
`potential-subtracted' (PS) mass scheme~\cite{Beneke:1998rk}.
While  conversion of the PS mass to the $\overline{\text{MS}}$ mass
scheme is a subject to additional theoretical uncertainties, this is
beyond the scope of this paper.


\section{Fit procedure}

\subsection{Simulated experiments}

  For each simulation of the threshold scan measurement, 
  one of the cross section templates (convoluted with the luminosity
  spectra) is selected as the base for pseudo-experiment generation. 
  By this choice, ``true'' values of top quark parameters are chosen
  for the considered fit scenario.
  Unless specified otherwise, nominal parameter values (and nominal
  template) were used in all calculations. 
        
\begin{figure}[tb]
  \centering
  \includegraphics[height = 6cm]{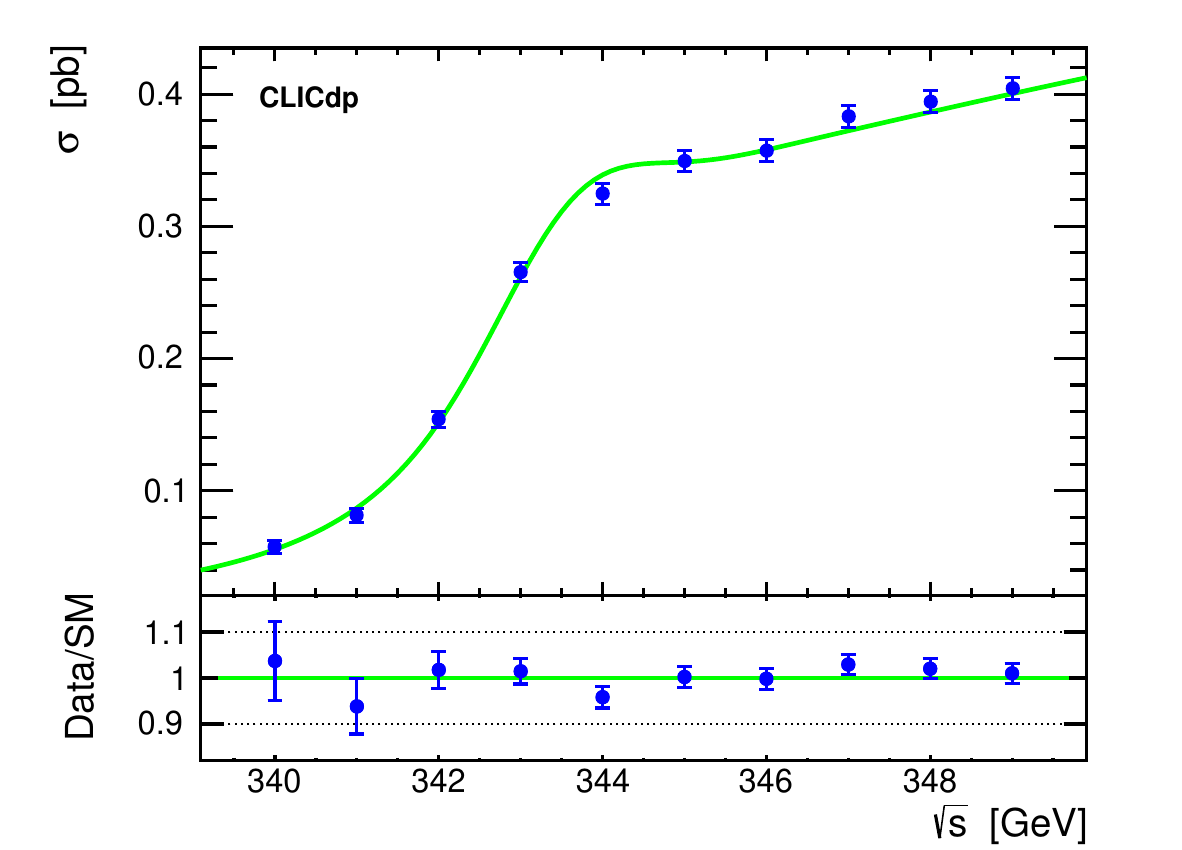}
  \caption{Reference scenario for the top-quark threshold scan. The
    error bars on the simulated data points show the expected
    statistical uncertainties of the cross section measurements
    assuming integrated luminosity of 10\,fb$^{-1}$ per
    measurement. 
    \label{fig:losowanie}
    }
   \end{figure}

  Selected template is used to generate a set of data corresponding to
  the expected scan results at CLIC, so called pseudo-experiment data
  (or pseudo-data).  
  For each scan point, cross section value taken from the base
  template is varied according to the expected statistical
  fluctuations (from Poisson distribution), taking selection
  efficiency and background contribution into account.
  Example of the generated pseudo-data set is shown in
  Fig.~\ref{fig:losowanie}. 

\subsection{Minimisation method}
         \label{sec:fit_method}
         
 Parameter fit procedure is then applied to the generated pseudo-data
 set. For each considered cross section template the $\chi^2$ value is
 calculated from the formula
 \begin{equation}
     \chi^2(\vec{p},\alpha)  = 
              \sum_{i} \left( \frac{m_i - \alpha\cdot \mu_i(\vec{p})}{\sigma_i} \right)^2
              \; + \; \left(\frac{\alpha - 1}{\Delta}\right)^2
              \; + \; \sum_j \left(\frac{p_j - \tilde{p}_j}{\sigma_{p_j}}\right)^2
              \label{eq:chi}
 \end{equation}
 where $m_i$ and $\sigma_i$ are the measured cross section values with
 their statistical uncertainties (pseudo-data set), $\mu_i(\vec{p})$
 denotes the template  cross section values for corresponding
 collision energies and given parameter set $\vec{p} = (m_t, \Gamma_t,
 \alpha_s, y_t^2, f_{bg})$,  and $\alpha$ is the template
 normalisation factor.
 The second term in the $\chi^2$ formula corresponds to the
 normalisation constraint, where different values of the relative
 normalisation uncertainty $\Delta$ are considered.
 The normalisation factor $\alpha$ can be evaluated analytically
 separately for each template by solving the minimisation condition 
 $\frac{\partial\chi^2}{\partial\alpha}=0$.  
 The third term represents possible constraints on the model parameters
 $p_j$ resulting from the earlier, independent measurements 
 with uncertainty $\sigma_{p_j}$. 
 In this study, external constraints are considered
 for $\alpha_s$, $y_t$ and $f_{bg}$.
            
 The values of the top-quark mass and other model parameters can then
 be extracted from the fit of the polynomial $\chi^2$  dependence  
 on the components of $\vec{p}$
 \begin{equation}
   \chi^2_\alpha(\vec{p}) = \sum_{i = 0}^{N} \sum_{j = 0}^{N} A_{i,j}\; p_i p_j \label{eq:para}
 \end{equation}
 where $\chi^2_\alpha$ is the $\chi^2$ value minimised
 w.r.t. normalisation factor $\alpha$, $N$ is the number of considered
 model parameters and $i, j$ are parameter indexes ($i, j = 1\ldots
 N$). 
 $A_{i,j}$ is the symmetric matrix of coefficients of N-dimensional parabola.
 To simplify the formula, $p_0 \equiv 1$ was defined (the linear part
 is thus given by $2A_{0,j}p_j$ and the constant part is $A_{0,0}$).
 The values of  parameters $A_{0,0}, \ldots, A_{n,n}$ are found by
 solving a set of linear equations, where $\chi^2_\alpha$ values 
 calculated for different parameter sets $\vec{p}$ are used as input.
 The fitted values of $A_{0,0}, \ldots, A_{n,n}$ are then used to
 extract the parameter values minimising the $\chi^2_\alpha(\vec{p})$,
 which constitute the fit result.
 Statistical uncertainties of the fitted parameter values, 
 $\vec{\sigma}=(\sigma_{m_t},\sigma_{\Gamma_t},\sigma_{\alpha_s},\sigma_{y_t^2}, \sigma_{f_{bg}})$,
 and their correlation coefficients, $r_{i,j}$, 
 are extracted from the fit covariance matrix.
 In the following, average values of the parameter
 uncertainties resulting from a large number of fits to the equivalent
 pseudo-data sets, are quoted as the expected parameter uncertainties
 from the given fit configuration.

 \begin{figure}[tb]
     \centering
     \includegraphics[height = 6cm]{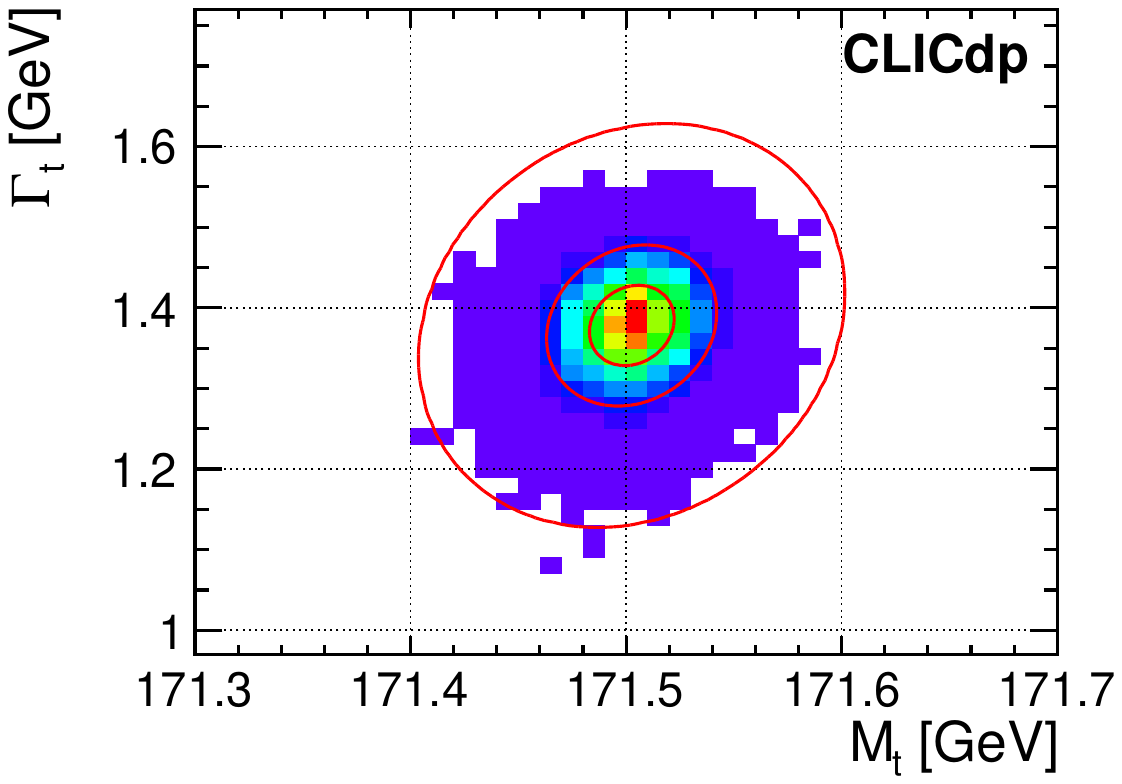}
     \caption{Results of the two-parameter fits to the top threshold
       scan, for sets of pseudo-data generated for nominal parameter
       values, $\tilde{m}_{t} = 171.5$~GeV and
       $\tilde{\Gamma}_{t}=1.37$~GeV.
       One, two and five sigma contours represent the 2-D Gaussian
       distribution fitted to the presented results.} 
     \label{fig:kontur}
 \end{figure}
The fitting procedure was applied to a large sample of pseudo-data
sets generated for the reference scan scenario (see
Fig. \ref{fig:losowanie}) and the nominal parameter values (see
Tab.~\ref{tab:qq_par}), taking top-quark mass and width as the only
free parameters in the fit. 
Distribution of the fit results presented in  Fig.~\ref{fig:kontur}
is well described by the 2-D Gaussian profile; results of the profile
fit to the data are indicated by the one, two and five sigma contours
indicated in the  plot. 
 For the considered 2-D fit configuration (top-quark mass vs top-quark
 width), fits to the marginalized parameter distributions result in $\sigma_{m_t}$=19.4\,MeV and
 $\sigma_{\Gamma_t}$=49.3\,MeV. 
 These uncertainties are in a good agreement with results of the previous
 study \cite{clic-top}, see Fig.~\ref{fig:lum} (right).
 The uncertainties estimated from parameter distributions are also in
 very good agreement with those extracted from the covariance matrix
 of the fit.


\section{Baseline threshold scan}

    \label{ch:baseline}
    
        \subsection{Comparison of fit configurations}
        
 As mentioned above, study \cite{clic-top} considered  
 two-parameter fits to the threshold scan data only. 
 This is also the case for earlier studies  \cite{Seidel,Horiguchi},
 while simultaneous fit of four model parameters was previously
 considered in \cite{Martinez:2002st}.
 Presented approach, thanks to its simple, semi-analytical form, allows to
 perform fits with even more free model parameters.
 Moreover, it is possible to add additional constraints on the
 selected parameters
 to make the fit reflect the expected experimental situation.

 \begin{figure}[tb]
     \centering
     \includegraphics[height = 6cm]{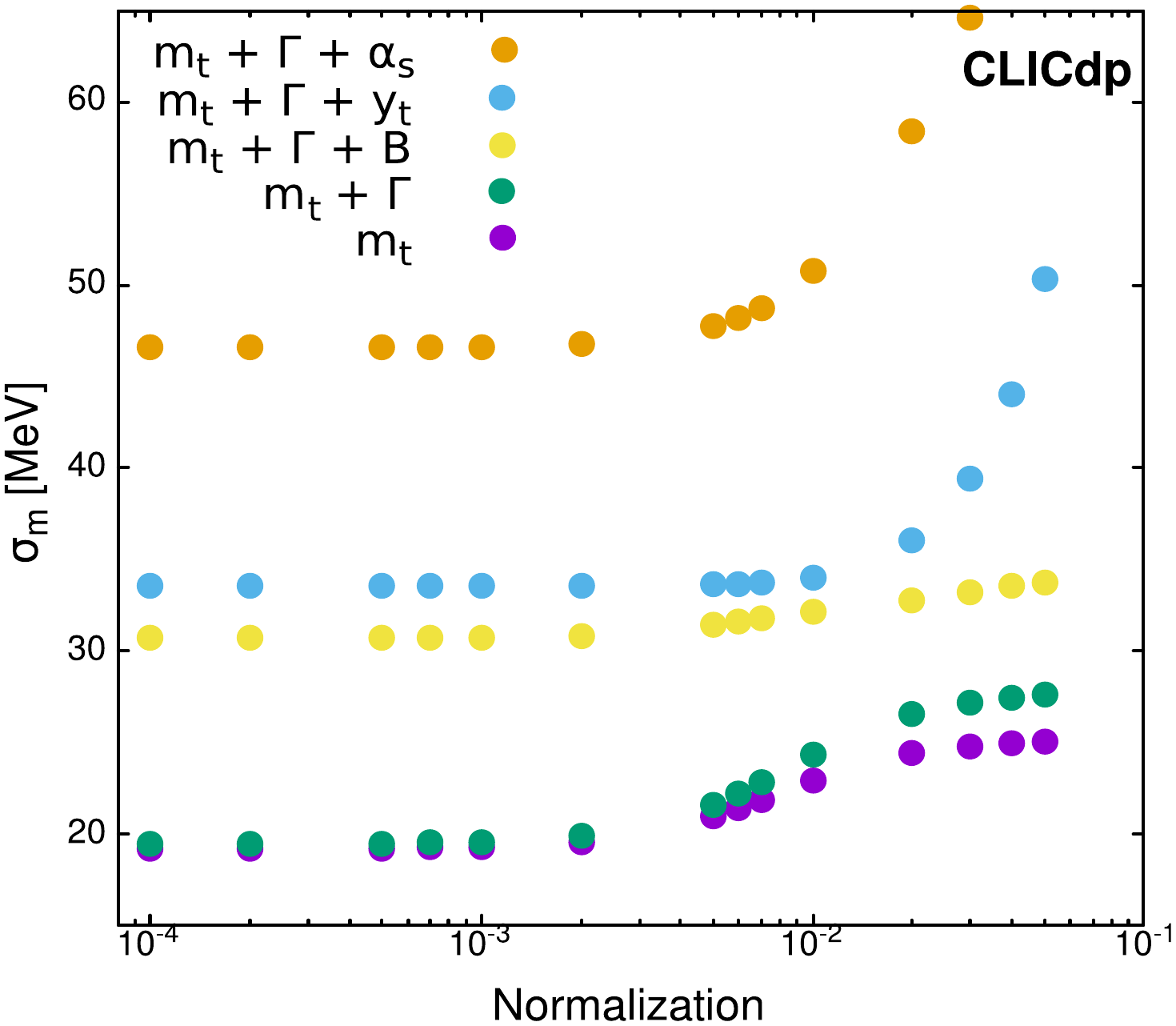}
     \includegraphics[height = 6cm]{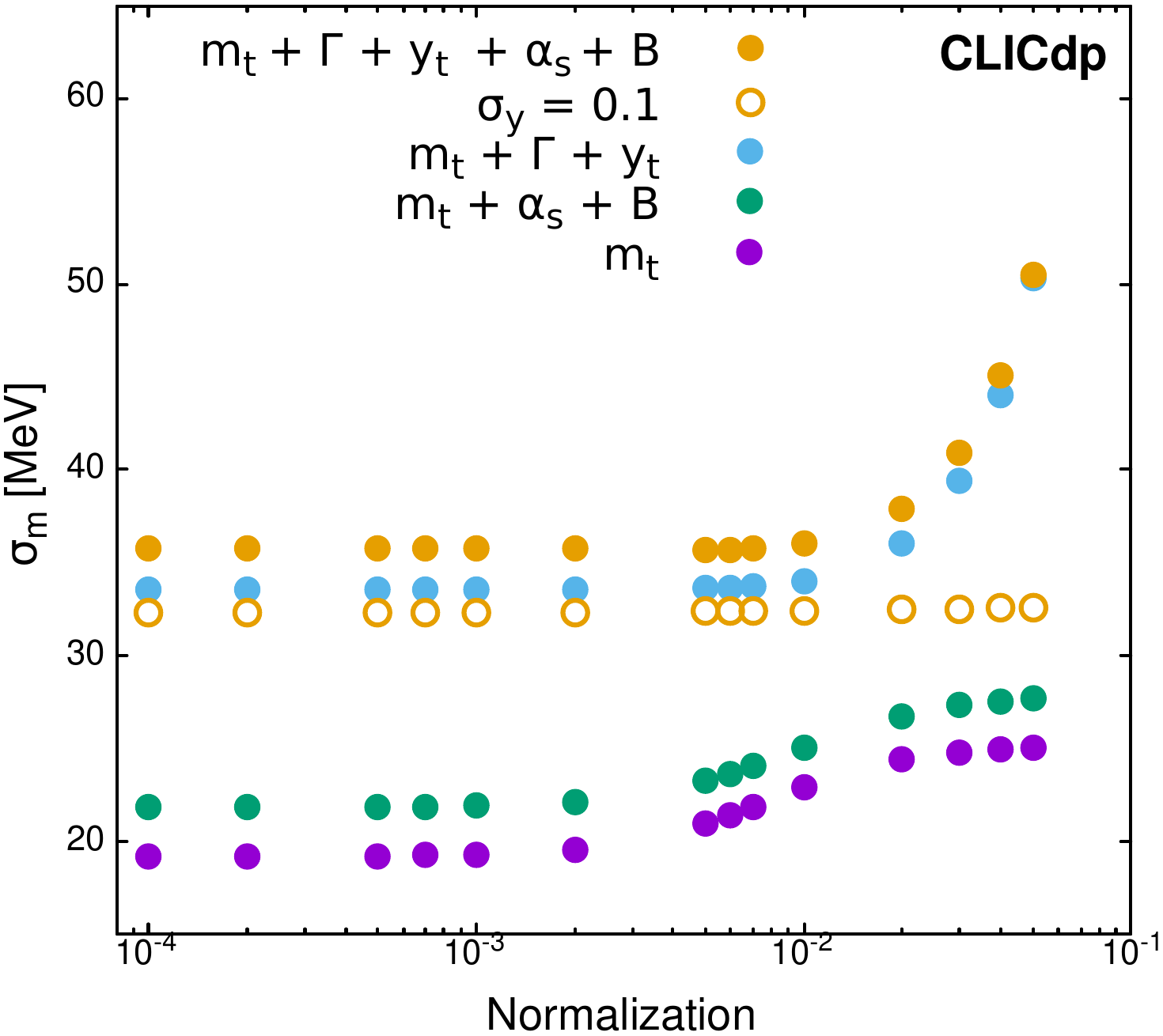}
     \caption{ Statistical uncertainty on the top-quark mass as a
       function of the assumed overall normalisation uncertainty for
       different fit configurations.
Left:
without any additional parameter constraints. 
Right: for ``SM constrained'' mass fit ($m_t$) and
the fit of mass, width and Yukawa coupling  ($m_t + \Gamma_t + y_t$),
without and with systematic parameters and their constraints included
in the fit; 
results of the fit with additional constraint on the Yukawa coupling
are also indicated, see text for details.} 
     \label{fig:norm4} 
 \end{figure} 
 Presented in Fig.~\ref{fig:norm4} (left) is
 the expected statistical precision of the top-quark mass
 as the function of the assumed normalisation uncertainty, $\Delta$, 
 for five selected configurations of the threshold scan data fit.
Precision of the one-parameter (1D) mass fit
($m_t$) is compared with the two-dimensional (2D) fit of mass and
width ($m_t + \Gamma_t$) and different three-dimensional
(3D) fit configurations.
Expected top-quark mass uncertainty from the 2D fit to the threshold
scan data is very close to the results of the 1D fit and sizeable
differences are only observed, if normalisation is not well
constrained, $\Delta > 0.01$. 
However, expected mass uncertainty increases significantly when adding
more (unconstrained) parameters to the fit.
Largest deterioration 
is observed when the strong coupling constant is considered as the
free model parameter ($m_t + \Gamma_t + \alpha_s$) with uncertainty
increasing by over a factor of two. 
Same fit configuration is also most sensitive to the
overall normalisation of the data.
This demontrates that there are significant correlations in the fit
between the top-quark mass and other model parameters (as well as the
data normalisation) and the external constraints on these parameters
are crucial for the precise top-quark mass determination.

In Fig.~\ref{fig:norm4} (right) top-qurk mass uncertainties expected
for two different fit approaches are compared. 
One-parameter fit of the top-quark mass ($m_t$) can be considered as a
model for ``SM constrained'' data analysis, when top-quark width and
Yukawa coupling  are taken from SM predictions.
When including systematic variations from the strong coupling
constant, $\alpha_{s}$, and the background level, $f_{bg}$,
uncertainty of the top-quark mass is increased by about 2.5\,MeV
independent on the assumed normalisation uncertainty ($m_t + \alpha_s + B$). 
With the normalisation constrained to better than 1\%,
uncertainty on the top-quark mass of below 25\,MeV is expected for
``SM constrained'' fit configuration.

Also presented in Fig.~\ref{fig:norm4} (right) are results of a more
general fit approach, when no relation is assumed between the
top-quark mass, its width and Yukawa coupling, allowing for possible
BSM contributions. 
For the three-dimensional fit of mass, width and Yukawa coupling ($m_t
+ \Gamma_t + y_t$), expected precision of the mass determination is
about 34\,MeV, assuming the normalisation is known to better than 1\%.
After adding the  two systematic parameters  to the fit,
the uncertainty increases by about 2\,MeV.
Independent measurement of the top Yukawa coupling at
CLIC~\cite{deBlas:2018mhx} or by other
experiments~\cite{Cepeda:2019klc} can improve the precision of the
top-quark mass determination to around 32\,MeV, assuming
$\sigma_{y_t}=0.1$. 
Constrain on the top Yukawa coupling reduces also the sensitivity to
the data normalisation, which can be attributed to the significant
correlations between the parameters describing the Yukawa coupling,
background level and the strong coupling constant. 

 \subsection{Impact of constraints}
        
 As demonstrated above,  constraints on model parameter 
 resulting from measurements preceding the top threshold scan at CLIC
 can significantly reduce statistical uncertainties of the fit. 
 In this section, impact of these constraints is studied in a more
 quantitative way. 
 The problem can also be reversed: how precisely should other
 model parameters be measured in order to allow for the best possible
 top-quark mass determination in the threshold scan.
 Considered is the most general fit configuration with all model
 parameters included in the five-dimensional (5D) fit: top-quark
 mass, $m_t$, width, $\Gamma_t$, and the Yukawa coupling,
 $y_t$, as well ass the strong coupling constant, $\alpha_s$ and the
 background level scaling factor, $f_{bg}$ (see Tab.~\ref{tab:qq_par}).
 Note that the data normalisation factor, $\alpha$, the sixth model
 parameter, is not included in the fit procedure but is evaluated
 analytically for each cross section template, as 
 described in Sec.~\ref{sec:fit_method}.

 \begin{figure}[bt]
     \centering
     \includegraphics[height=6cm]{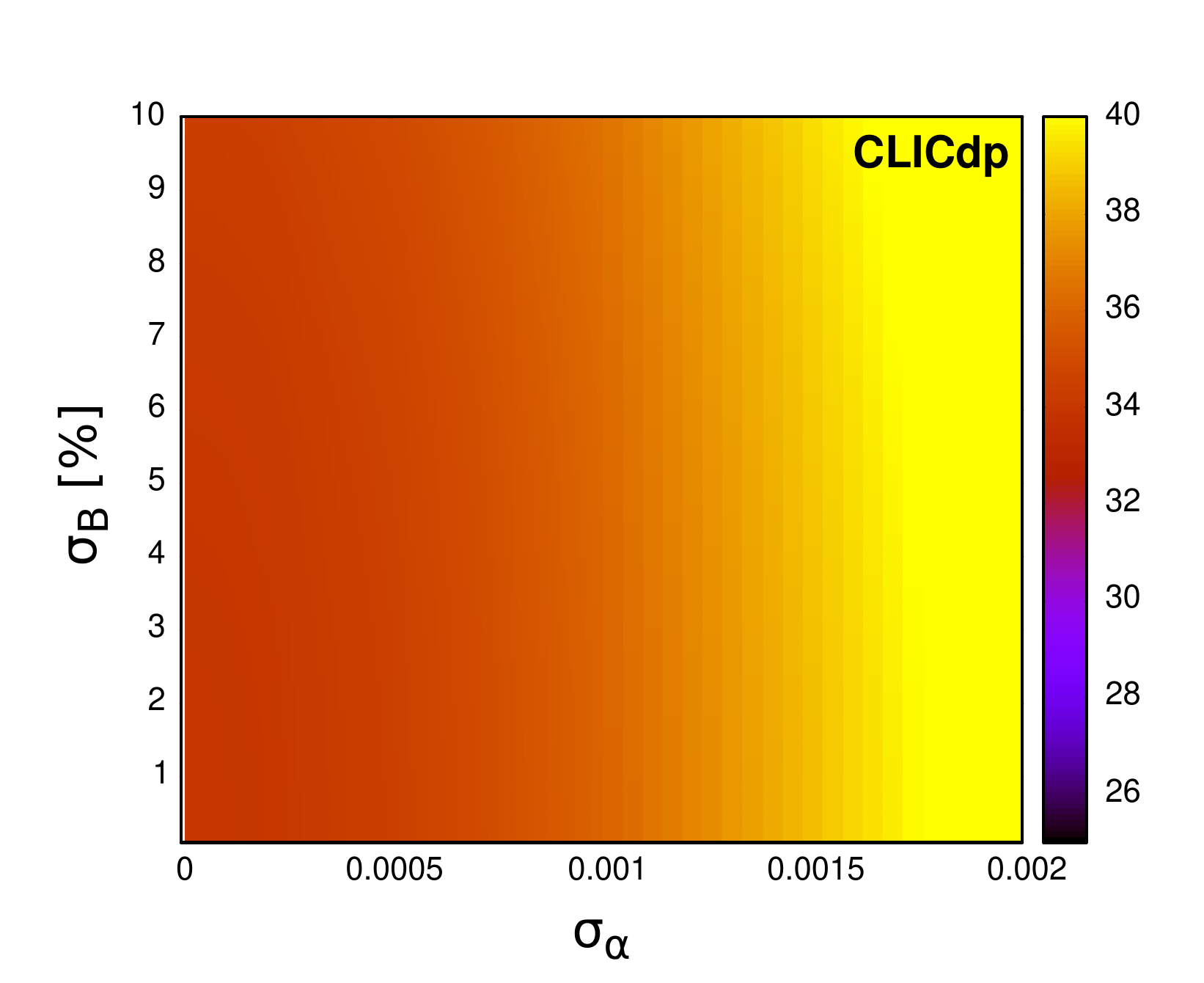}
     \includegraphics[height=6cm]{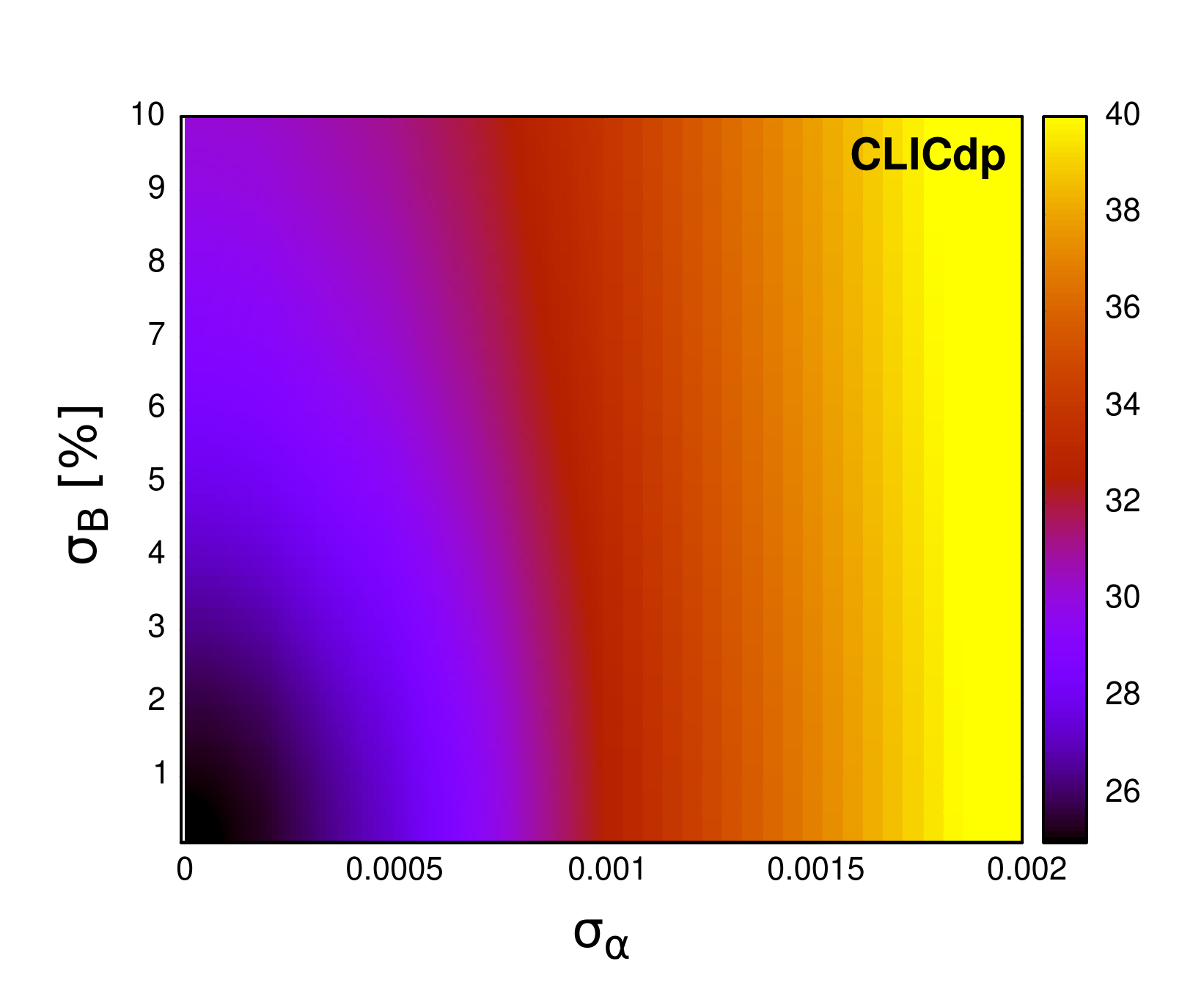}
     \caption{ Expected uncertainty on the top-quark mass
       as a function of the uncertainty of the strong coupling
       constant and the background level uncertainty. Results from the
       fit of all model parameters (5D) are presented with  
     no constraint on the top Yukawa coupling (left) and for the
     assumed Yukawa uncertainty of $\sigma_{y_t} = 0.1$ (right). 
     Normalisation uncertainty, $\Delta = 1\%$, is assumed.
     Colour maps are obtained using a linear interpolation algorithm.}
     \label{fig:map3d}
 \end{figure}
 Shown in Fig.~\ref{fig:map3d} is the expected top-quark mass
 uncertainty from the 5D fit to the threshold scan data
 plotted as a function of the assumed strong
 coupling constant and background contribution uncertainties.
 Relative uncertainty of 1\% is assumed for the overall data normalisation.
 Two fit scenarios are considered: with the top Yukawa coupling considered
 an unconstrained model parameter (left plot) and with assumed Yukawa
 uncertainty from earlier measurements of $\sigma_{y_t}= 0.1$ (right). 
 If no constraint can be set on the value of the Yukawa coupling,
 uncertainty of the extracted top-quark mass depends mainly on the 
 assumed uncertainty on $\alpha_s$.
 Even if the strong coupling constant is known with very high
 precision, uncertainty on the mass cannot be reduced below about 34\,MeV
 (see also 3D fit results shown in Fig.~\ref{fig:norm4}).
 The estimated top-quark mass uncertainty decreases significantly,
 if external constraint on the Yukawa coupling, with $\sigma_{y_t}= 0.1$, 
 is imposed, as shown in  Fig.~\ref{fig:map3d} (right).
 However, to be able to extract top-quark mass with precision of
 the order of 25\,MeV, strong coupling constant would need to be known
 with precision higher than 0.0003 and the background contribution to
 at least  2\%.
 With the current uncertainty of the world average,  $\sigma_{\alpha_s} =
0.001$ \cite{pdg}, mass uncertainty cannot be reduced below about
32\,MeV.
Unfortunately, uncertainty on the strong coupling constant is not
likely to be significantly reduced in the near future~\cite{Strategy:2019vxc}.

\begin{figure}[tb]
     \centering
         \includegraphics[height=5.9cm]{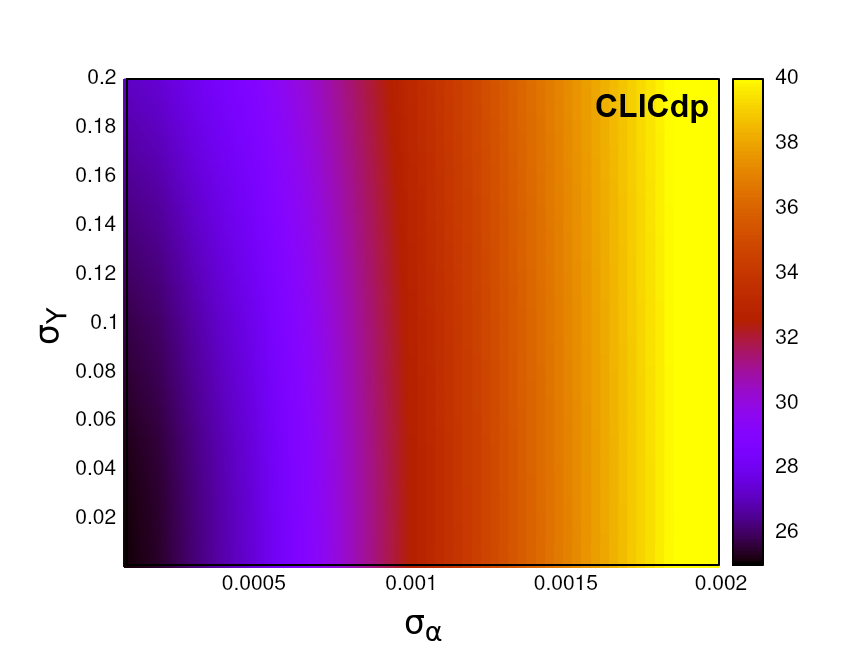}
         \includegraphics[height=5.9cm]{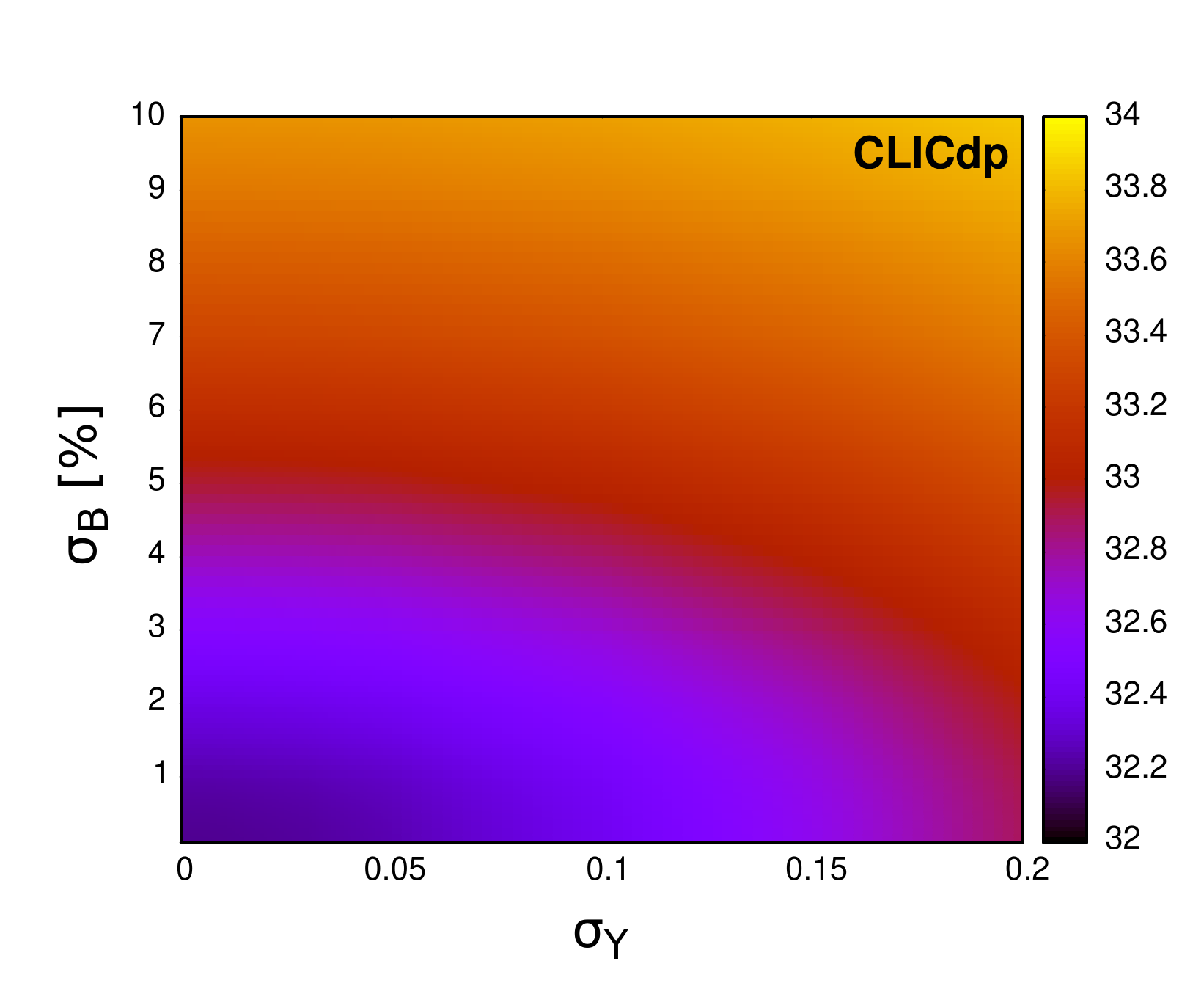}
     \caption{ Expected uncertainty on the top-quark mass  from the fit involving all
  considered model parameters (5D). 
  Left: as a function of the strong coupling constant and Yukawa coupling
  uncertainties, assuming background uncertainty of 2\% and  normalisation uncertainty, $\Delta = 1\%$. 
  Right: as a function of the background and Yukawa coupling uncertainties
  assuming strong coupling uncertainty of 0.001 and
  normalisation uncertainty, $\Delta = 1\%$. }
     \label{fig:map4d}
 \end{figure}

Impact of the assumed Yukawa coupling constraint on the expected fit
precision is also shown in Fig.~\ref{fig:map4d}.
Uncertainty on the top-quark mass is presented as a function of the
assumed Yukawa coupling and  the strong coupling constant
uncertainties (left plot), and as a function of the Yukawa coupling
and the background contribution uncertainties (right plot).
Presented results indicate that, although the extraction of
top-quark mass does profit from additional Yukawa coupling constraint,
possible improvement of the coupling determination precision beyond
the assumed uncertainty of $\sigma_{y_t}=0.1$, will hardly improve the
mass determination precision.  
Therefore this value of uncertainty is used for the following
analysis,
although higher precision is expected considering combined analysis of
the Higgs boson measurements at the HL-LHC \cite{Cepeda:2019klc}.

\begin{figure}[t]
     \centering
     \includegraphics[height=6cm]{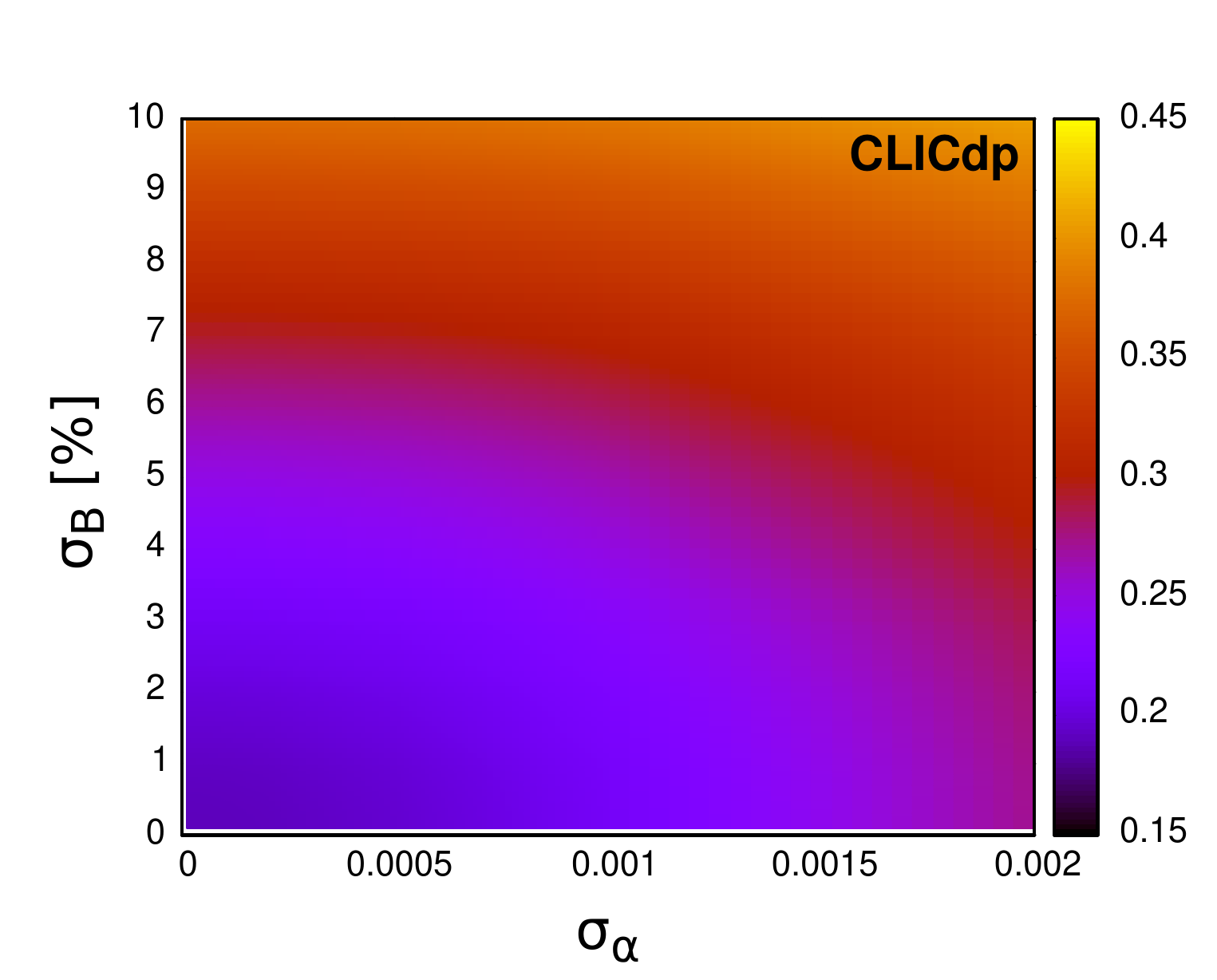}
     \caption{ Expected statistical uncertainty on the top Yukawa
       coupling  from the fit involving all considered model
       parameters (5D), as a function of the strong coupling constant
       and background contribution uncertainties.
     Normalisation uncertainty, $\Delta = 1\%$, is assumed.
     }
     \label{fig:ymap}
 \end{figure}

  As the top-quark pair-production cross section depends on the top
 Yukawa coupling, the threshold scan data can also be used to constrain
 its value.  
 The statistical precision of the Yukawa coupling
 determination from the 5D fit was investigated as a function of the assumed
 background and strong coupling uncertainties. 
 Results presented in Fig.~\ref{fig:ymap} indicate that contribution
 from the Yukawa coupling can be observed in the threshold scan data
 with 5$\sigma$ significance (i.e. with statistical precision of about 0.2)
 assuming the normalisation is known at percent level, the strong
 coupling is known better than 0.001 and the background uncertainty
 is below 3\%.

 \subsection{Systematic effects}

 Results presented above demonstrate that systematic uncertainties are
 likely to limit the ultimate precision of the top-quark mass
 determination from the threshold scan. 
 Various sources of uncertainties have been
 investigated in \cite{clic-top} and the combined systematic
 uncertainty of the top-quark mass is expected to be in range of
 30\,MeV to 50\,MeV.
 In particular, the parametric uncertainty originating from the strong
 coupling constant was estimated to be 30\,MeV (for the reduced charge luminosity
 spectrum considered here), also assuming an uncertainty of 0.001
 in $\alpha_{s}$.
 Systematic mass uncertainty resulting from the background level estimates
 is much smaller: for the 5\% shift of the estimated background
 contribution 18\,MeV variation is expected in the extracted
 top-quark mass \cite{Seidel}, corresponding to about 7\,MeV variation for
 the background level uncertainty of 2\% assumed here.

 The procedure described in this work, including parameters
 describing systematic effects and the corresponding constraints
 directly in the fit to the threshold scan data, allows for proper
 treatment of all uncertainties, including their correlations,
 resulting in a significant reduction of the total mass uncertainty.
 When systematic variations of the strong coupling constant and the
 background level are not considered, the expected statistical
 uncertainty of the top quark mass from the 3D fit is 34\,MeV
 (22\,MeV) without (with) additional constraint on the Yukawa
 coupling.
 With parametric uncertainties from $\alpha_{s}$ and $f_{bg}$
 variations of 30\,MeV and 7\,MeV, respectively, this would correspond
 to the total top-quark mass uncertainty of 46\,MeV (38\,MeV).
 With the proposed approach, where the threshold scan data can be also
 used to reduce the influence of the parameter variations, the
 final uncertainties (including the considered systematic effects) are
 36\,MeV (32\,MeV), see Fig.~\ref{fig:norm4}. This corresponds to the
 reduction of the total uncertainty by 15 to 20\%.


    \section{Scan optimisation}
    
        \subsection{Multi-objective optimisation}
        
When looking for the best scenario of the top-quark threshold
scan at CLIC (or at any other future \epem collider) one needs to
consider different aspects of the measurement.  
The top quark mass is not the only parameter that needs to be extracted
(with the best possible statistical uncertainty). There are other 
model parameters, measurement of which needs to be optimised at the same time.
Multi-objective optimisation problems are likely to be complex and
must be approached differently, depending on the specific case.   
When looking for the best optimisation approach it is necessary to take
into account, for example, how easy it is to find solution to the
considered problem  and how many solutions are expected.
It could be that there are so many possible solutions that all of them cannot
be analyzed.   
            
The easiest case is when one can propose a set of equations describing
the problem, defining the variables to be optimised.
These equations can be then solved analytical or numerically, to find the optimum. 
Yet, it is rarely the case when the problem can be reduced to such a
set of equations. 

In most real-world problems, heuristic procedures are being used to
find optimal solutions.
They include Genetic algorithms, that are inspired by biological
evolution processes, such as reproduction, mutation, recombination, and
selection \cite{evo}.
In a Genetic algorithm, a set of proposed solutions to an optimisation problem, 
called \textdef{Individuals}, is evolved towards better solutions. 
Each Individual has a defined set of properties, called
\textdef{genotype},  which can be mutated and altered, and set of
measurable traits, called \textdef{phenotype}, that are determined by
the genotype.
Phenotype consists of traits that are used to evaluate their
performance and choose the best Individuals to the next generation. 
During consecutive iterations of the algorithm population evolves
towards better solutions. 
Ultimately, after a finite number of iterations, population should
converge to the optimal solution. 

\subsection{Non dominated sorting genetic algorithm II}
        
From many available Genetic algorithms, Non dominated
sorting genetic algorithm II, proposed by Kalyanmoy Deb in 2002
\cite{deb}, was chosen for this study, for its high efficiency and elasticity.
Each iteration of the algorithm is divided into three steps: 
\begin{itemize}
  \setlength{\itemsep}{0pt} 
    \item Creating Children and adding them into population
    \item Non dominated sorting of the population (with Jensen
      algorithm \cite{jensen}) 
    \item Choosing the best Individuals for the next generation
\end{itemize}
Each one of those steps is highly customizable, so it is possible to adjust the
implementation in order to achieve the best results. 
Thanks to use of efficient non dominated sorting algorithm it was possible to
lower time complexity from  original $O(MN^2)$ \cite{deb} to $O(MN
log^{M-1}N)$ \cite{jensen}, where $M$ is number of objectives and $N$
is the size of the population.

\subsubsection*{Creating Individuals and First Generation}

The first step was to translate our measurement procedure into Genetic
algorithm language.  
One measurement scenario was assumed as an Individual, which genotype is
represented by a scan sequence, set of centre-of-mass energy points 
(scan points make a multiset, as they can repeat, but their order is
irrelevant).
In this way each scan point can be considered a chromosome.
Constant total integrated luminosity of 100\,fb$^{-1}$ was assumed
and its equal sharing among all scan points.
This might seem restrictive, but scan points with higher luminosity 
are actually allowed by allowing energy points to repeat in the scan sequence.
This way it is also possible to optimise luminosity distribution.  
The measurement procedure described in previous chapters translates
scan sequence (genotype) into measured top quark parameters with their
uncertainties (phenotype).
Yet, this process is non-deterministic, as statistical fluctuations 
were drawn from Poisson distribution separately for each generated
pseudo-data set.
To take statistical fluctuations into account pseudo-experiment procedure 
is always repeated three times and the worst result for
each objective is taken.\footnote{The number of pseudo-experiments run for each
  Individual is limited by the processing time. However, already with
  three experiments the possible bias of the optimisation due to
  statistical fluctuations is basically excluded.}  
The First Generation is seeded by creating 2000 identical Individuals
using baseline scenario. Results presented in Chapter~\ref{ch:baseline} show
that it provides good fit results for all parameters.  
    
\subsubsection*{Pairing and breeding Individuals}

There are many proposed strategies to find the best pair of parents,
but they are computationally costly.
In order to ensure diversity in population, each Individual has three children,
one with each of three other, randomly chosen Individuals.
This method was used because of computational efficiency and for
maximizing diversity in population. 
After obtaining two parent genotypes, it is necessary to ensure they are of the
same length before making a new one (a child).
It was done by inserting empty chromosomes into a shorter one, to mach
their sizes.
While iterating over parental genotypes, parent chromosome to insert 
into child's genotype is chosen randomly.
To allow changing genotype length, 5\% chance to drop any of 
the chromosomes was added.
To avoid systematic loss of chromosomes, additional random
chromosome was added at the end of the genotype with a 10\% probability.
It was always required that the length of the resulting genotype is not
shorter than 2, as shorter ones cannot be used for the phenotype
evaluation (threshold fit procedure).
When copying parent chromosome, it is also shifted by a random mutation in a given
range. At the beginning of the evolution, the mutation range is $\pm
0.5$\,GeV, but it shrinks geometrically in each iteration by factor of
0.9. In this way, it was possible to quickly mutate the initial baseline scenario
without loosing the ability to converge around the best solution after
a larger number of iterations.  After creating all children in the
generation their phenotypes were computed and they were inserted into the
population. 
    
\subsubsection*{Performance evaluation and selecting the next Generation}

When considering more than one objective in the scan optimisation, one
cannot simply compare (and sort) Individuals by their phenotype
(parameter uncertainties from the fit).
Instead, desired configuration is Pareto efficiency, which is defined as a
configuration that cannot be modified so as to improve any objective 
without making at least one other objective worsen \cite{pareto}.
In order to find optimal solution Individuals must be compared with
each other.
To find which one is better, their Pareto dominance must be checked, which is
the situation when one Individual is better in at least one objective,
and not worse in all other \cite{pareto}.
The condition for Individual $x$ dominating Individual $y$,
$x \succ y$, can be described by the formula:
    \begin{equation}
        x \succ y \iff \forall_i : f_i(x) \geq f_i(y) \land \exists_j : f_j(x) > f_j(y)
    \end{equation}{}
where $f_i(x)$ denotes objective $i$ calculated for Individual $x$. 
When neither of them dominate the other, then the two Individuals are
Pareto efficient.
Individuals were grouped based on this criteria and such groups are
called Pareto Frontiers \cite{pareto}. They can by described by the
formula: 
    \begin{equation}
      P(Y) = \{y' \in Y : \{y'' \in Y : y'' \succ y'
      \} = \emptyset \} 
    \end{equation}{}
    Individuals were chosen for the next generation by sorting Pareto frontiers. It is based on criteria 
    that a solution in $Front_{k+1}$ must be dominated by at least one solution in $Front_k$ and may or may not
    dominate solutions in $Front_{k+2}$ \cite{fang}. To efficiently perform those calculations Jensen 
    algorithm  was used~\cite{jensen}.
    After sorting Pareto frontiers, 2000 Individuals 
    from frontiers with lowest ranks were selected for the next generation.


\section{Results}
\label{sec:results}

\subsection{Single objective optimisation}

The performance of the optimisation procedure based on the Genetic
algorithm was first studied for the optimised measurement of one model parameter
only: top-quark mass, top-quark width or top Yukawa coupling.
The objective is to minimize the statistical uncertainty resulting
from the fit of the given parameter in a most general (5D) fit procedure,  
with all model parameters taken into account.
Population size is set to 2000 and number of generations to 30. All
results presented in this section were calculated assuming
normalisation uncertainty of $\Delta=0.1\%$, strong coupling constant
uncertainty of $\sigma_{\alpha_s}=0.001$ and background level
uncertainty of $\sigma_{f_{bg}}=2\%$.
When measurement of the top Yukawa coupling was not included in the
optimisation objectives, uncertainty of $\sigma_{y_t} = 0.1$ was assumed
for constraint coming from independent coupling measurement.

\begin{figure}[tb]
\centering
\includegraphics[height=5.9cm]{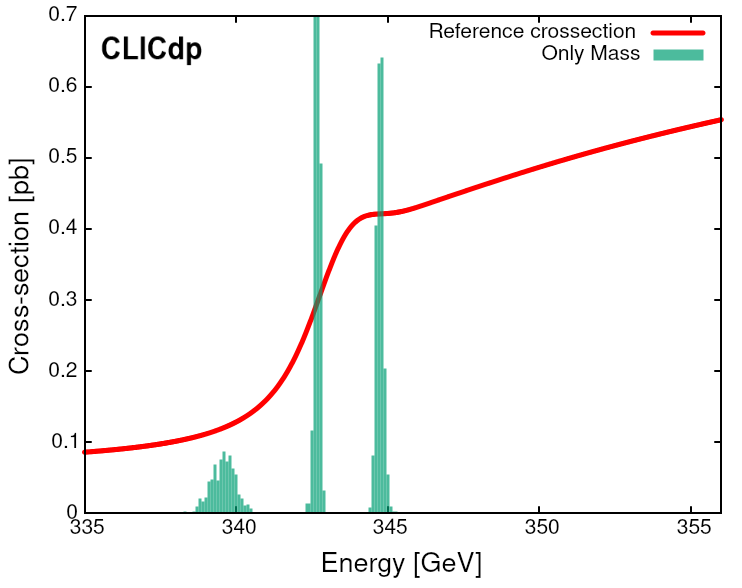}
\includegraphics[height=5.9cm]{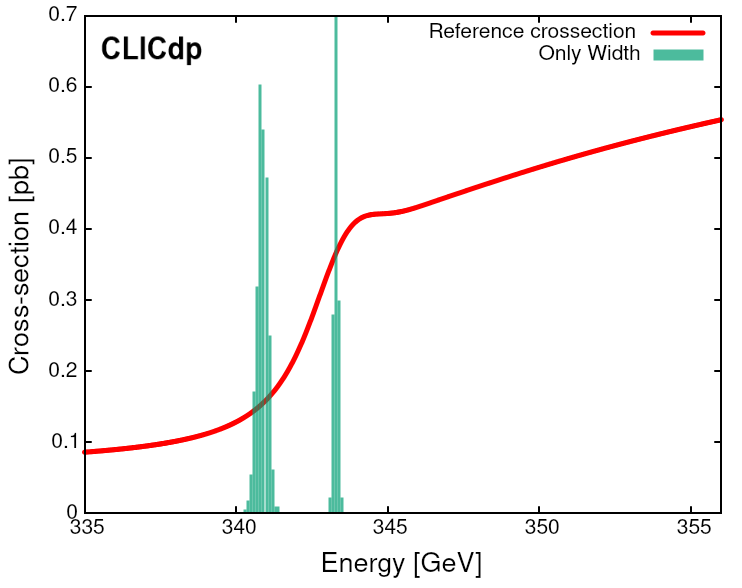}
\includegraphics[height=5.9cm]{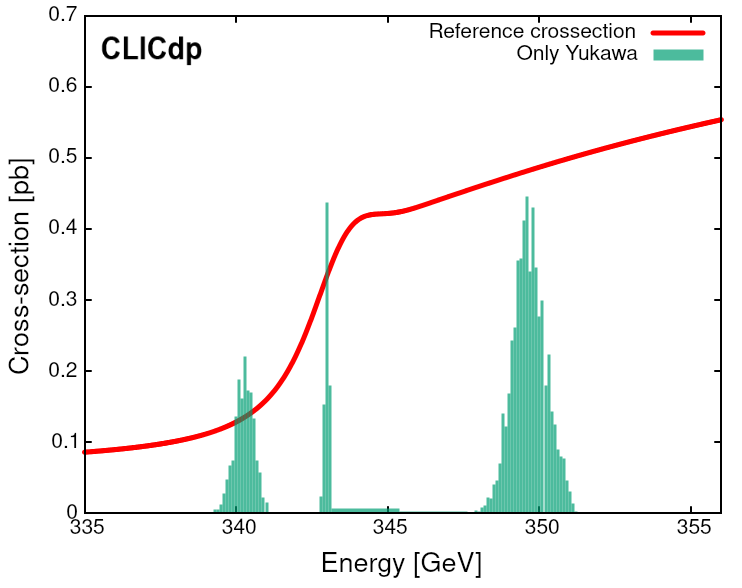}
 \caption{Distribution of the measurement points from last generation (arbitrary scale)
     compared with the reference cross section template, for three single parameter optimisation scenarios: top-quak mass (top left), top-quark width (top right) and top Yukawa coupling (bottom plot) measurement optimisation. 
   }
 \label{fig:single3}
 \end{figure}

Results of the optimisation procedure for the three considered model
parameters are summarised in Fig.~\ref{fig:single3}. 
When the scan procedure is optimised for the top-quark mass determination, 
the measurement is focused in three energy regions, with most of
the luminosity taken in the middle of the threshold (around 343\,GeV),
where the cross section  slope is highest, and at the plateau just
above the threshold (around 345\,GeV), see top left plot in Fig.~\ref{fig:single3}.  
With this choice of measurement points, it is possible to reduce the
expected uncertainty of the top-quark mass by about 30\%, from 
32\,MeV expected for the ten-point baseline scan  scenario to 22\,MeV
for the optimised ones.

However, the scenario optimised for mass measurement is clearly not
the best one when the top-quark width measurement is considered.
In this case, much larger fraction of the luminosity should be devoted
to the cross section measurement just below the threshold (around
341\,GeV),  see top right plot in Fig.~\ref{fig:single3}. 
With the width-optimised scan scenario, the statistical uncertainty on the
top-quark width is reduced by about 40\%, from 58\,MeV for the baseline scenario 
to around 35\,MeV for the optimal ones.

Smallest improvement due to scan optimisation can be obtained for the 
top Yukawa coupling measurement.
As shown in Fig.~\ref{fig:single3} (bottom plot), in order to
constrain the Yukawa coupling better, 
significant fraction of the luminosity has to be taken above the
threshold, at around 350\,GeV. 
This allows for about 20\% improvement in the precision of the top
Yukawa coupling determination, from about 0.18 to 0.14.

Looking at the measurement point distributions in Fig.~\ref{fig:single3} one
can clearly see that four energy regions are relevant in the scan
optimisation: energies just below the threshold (around 340-341\,GeV), in
the middle of the threshold (around 343\,GeV), on the plateau just
above the threshold (around 345\,GeV) and further above the threshold
(around 349-350\,GeV).
Different regions are sensitive to different model parameters:
the regions in the middle and just above the
threshold are most relevant for the  mass measurements, 
measurement below the threshold is crucial for the width measurement,
whereas the Yukawa coupling determination depends on the amount of
luminosity which can be devoted to running few GeV above the threshold.
This comparison shows that the same choice of energy points can be
optimal (or close to optimal) for all considered parameters. 
However, sharing of the luminosity between four energy regions depends
on the optimisation goal.

\subsection{Multiple objectives optimisation} 

When considering multiple objectives, it was decided to focus on pairs of top-quark parameters, 
in order to study how they influence each other in the optimisation procedure. 
\begin{figure}[t]
        \centering
\includegraphics[height=5.7cm]{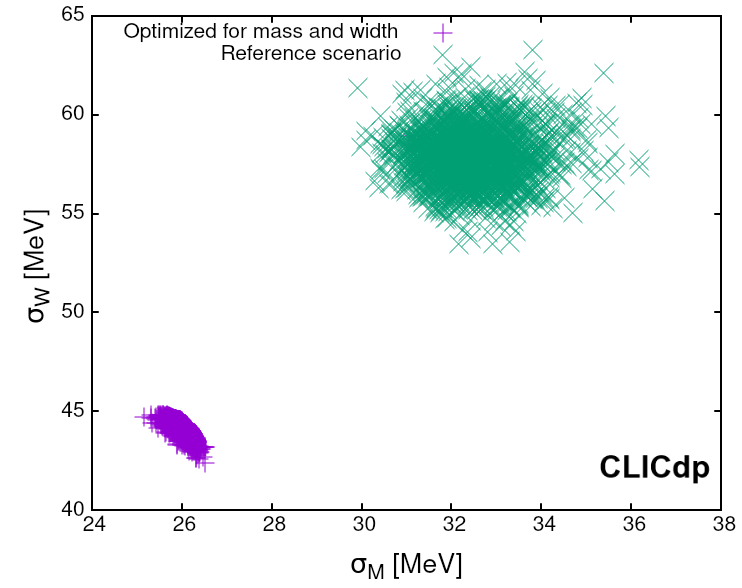}
\hspace{0.5cm}        
\includegraphics[height=5.7cm]{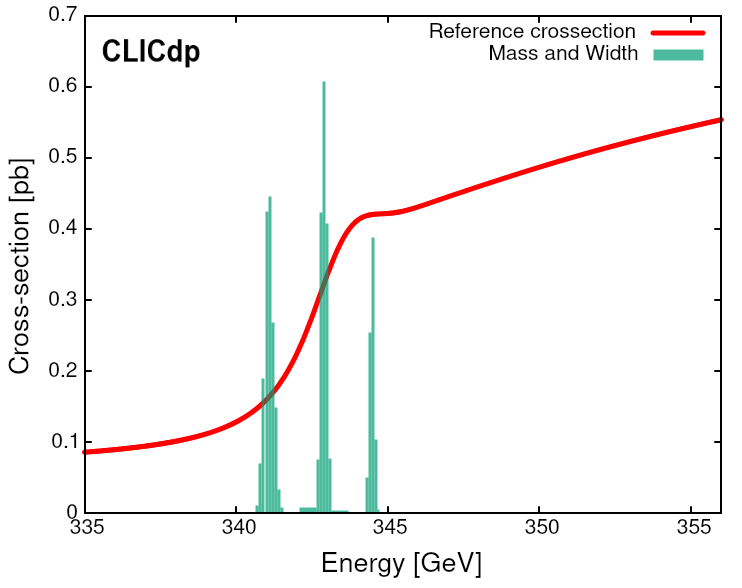}
        \caption{Left: mass and width uncertainty distribution in the first (green) 
        and the last (blue) generation for scan optimised for both mass and width determination precision. Right: distribution of the measurement points from the last generation (arbitrary scale) compared with the reference cross section template. }
        \label{fig:multiMW}
    \end{figure}
Shown in Fig.~\ref{fig:multiMW} are the results of mass and width measurement optimisation. 
For both parameters, improvement of about 20-25\% can be achieved:
mass uncertainty is reduced by about 6\,MeV, from 32\,MeV to 26\,MeV,
while the width uncertainty is reduced by about 14\,MeV, from 58\,MeV to 44\,MeV.
The improvement is smaller, when compared to single objective 
optimisation, by 10-15\% (about 4\,MeV for mass and 9\,MeV for width uncertainty).
As expected, the optimised measurement point distribution combines the
scenarios obtained for one parameter mass and width optimisations (see
Fig. \ref{fig:single3}).
Most of the scenarios from the last generation, more than 99\%,
include 5 energy points, showing good convergence of the
optimisation. 

\begin{figure}[t]
        \centering
\includegraphics[height=5.7cm]{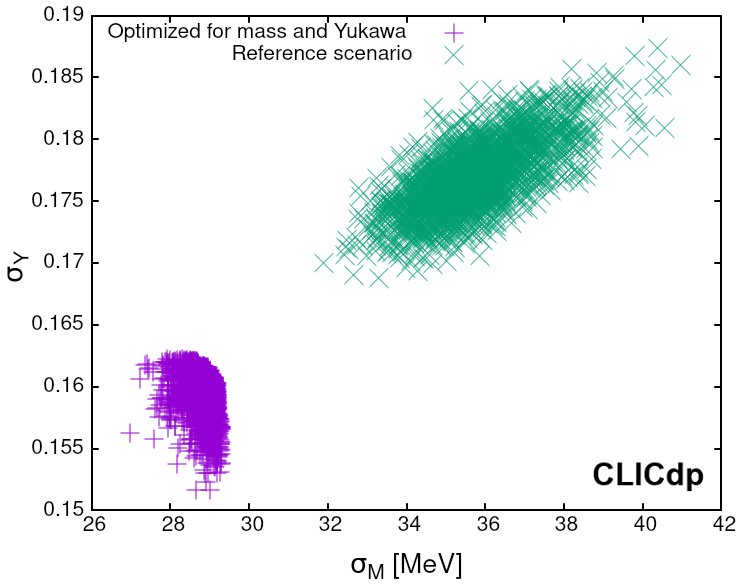}
\hspace{0.5cm}        
\includegraphics[height=5.7cm]{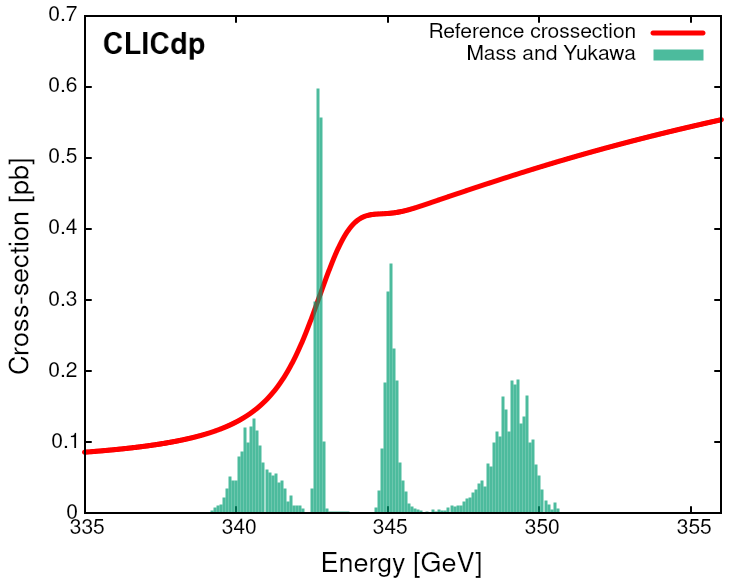}
        \caption{As in Fig.~\ref{fig:multiMW} but for scan optimised for mass and Yukawa coupling determination precision. }
        \label{fig:multiMY}
    \end{figure}

When optimising the scan procedure for simultaneous top-quark mass and
top Yukawa coupling measurement, see Fig.~\ref{fig:multiMY}, results
are again similar to those obtained for one objective
optimisation. 
Note that possible constraint on the Yukawa coupling is not taken into account
for this optimisation.
Improvement of about 20\% can be expected for the mass measurement, from 36\,MeV to 29\,MeV, but only
about 10\% improvement in the Yukawa coupling uncertainty. 
The measurement point distribution for this optimisation scenario is
very similar to the one obtained for Yukawa only optimisation (see
bottom plot in Fig. \ref{fig:single3}), but with additional
measurements just above the threshold (at about 345\,GeV) required for
precise mass determination.
More scan points are also required for the optimal measurement in this
case: 97\% of the last generation scenarios consist of 9 or 10
measurement points. 

\begin{figure}[t]
        \centering
\includegraphics[height=5.7cm]{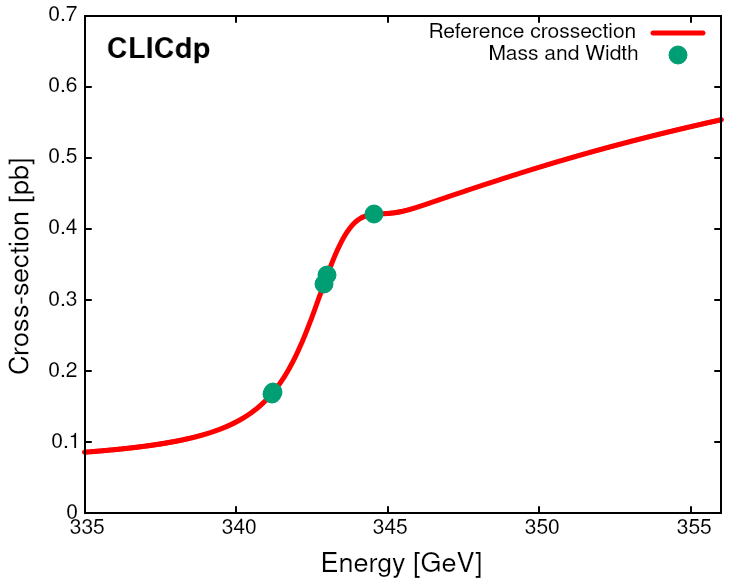}
\hspace{0.5cm}
\includegraphics[height=5.7cm]{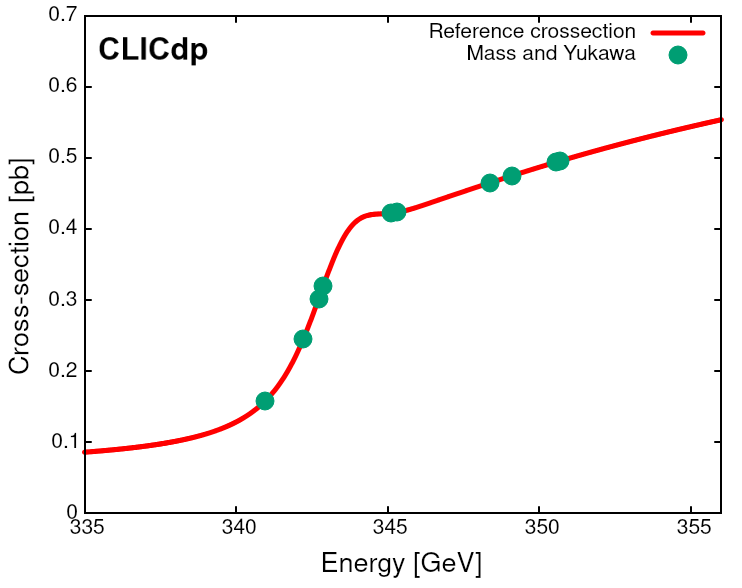}
        \caption{Scan energy points for the ``best scenarios'' taken from the last generation compared with 
        the reference cross section template: (left) 5 point scenario optimised for mass and width determination 
        precision (two points below, two in the middle and one above the threshold) and (right) 
        10 point scenario optimised for mass and Yukawa coupling determination precision. }
        \label{fig:scen}
    \end{figure}

For detailed comparison of the optimised scenarios with the baseline
threshold scan configuration, one scenario was selected from the last
generation for each of the aforementioned multi objective optimisations. 
Two selected scenarios are presented in Fig.~\ref{fig:scen}:
five-point scenario optimised for mass and width measurement 
and ten-point one, optimised for mass and Yukawa coupling. 
In both scenarios points group in pairs, in places where more
luminosity should be collected (at the threshold and, in case of mass
and Yukawa optimisation, just above the threshold).
These ``best scenarios'' were compared with the reference scenario
based on results coming from 20\,000 pseudo-experiments. 
As before, normalisation uncertainty is $\Delta=0.1\%$, strong coupling
uncertainty is $\sigma_{\alpha_{s}}=0.001$, background level
uncertainty is $\sigma_{f_{bg}}=2\%$ and, for mass and width
optimisation, Yukawa uncertainty of $\sigma_{y_t}= 0.1$ is assumed.
Uncertainty of the ``true'' top quark mass assumed when defining the scan
sequence was not included at this point, it will be discussed in the next
subsection. 

    \begin{figure}[t]
\centering
\includegraphics[height=5.7cm]{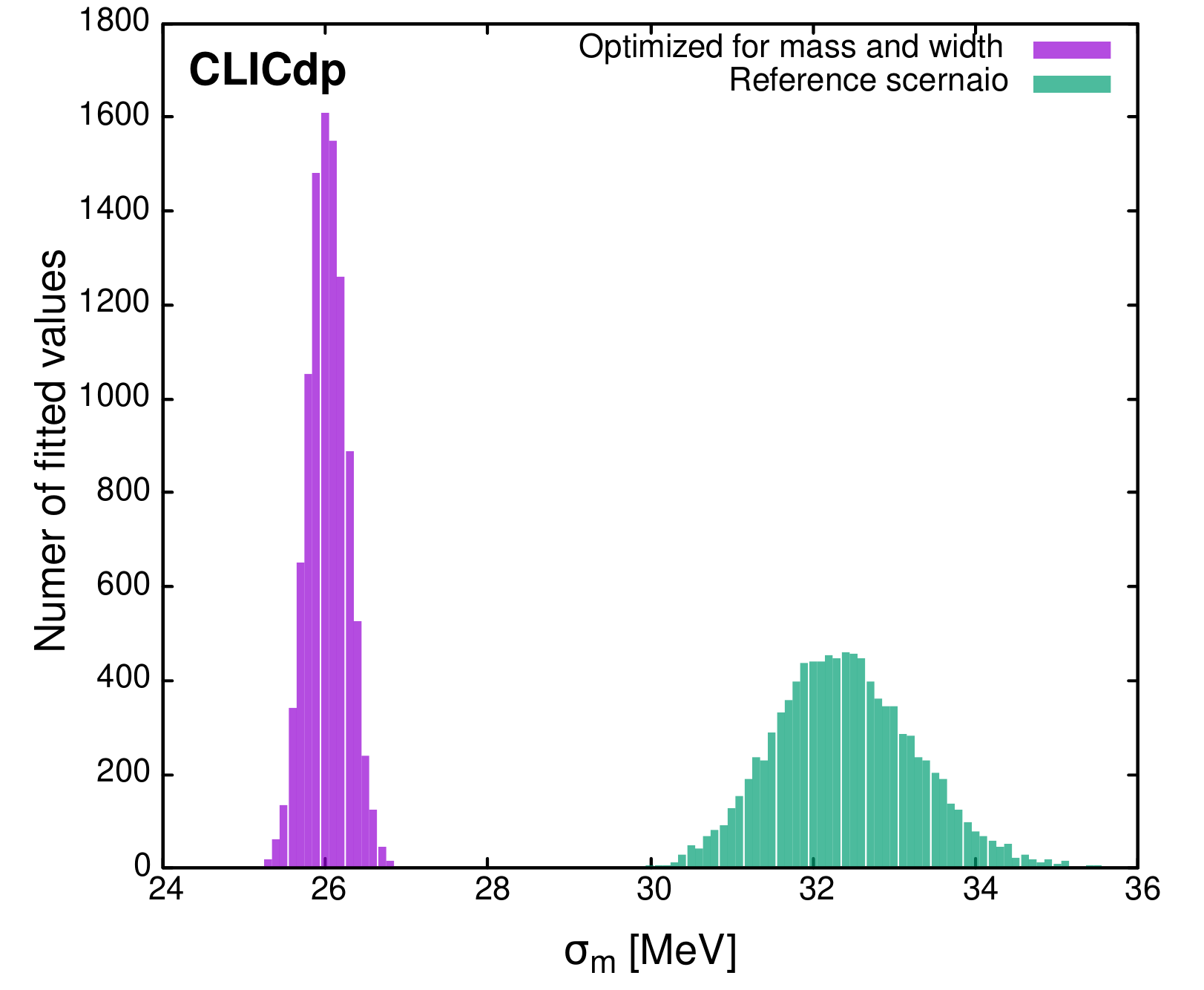}
\hspace{0.5cm}
\includegraphics[height=5.7cm]{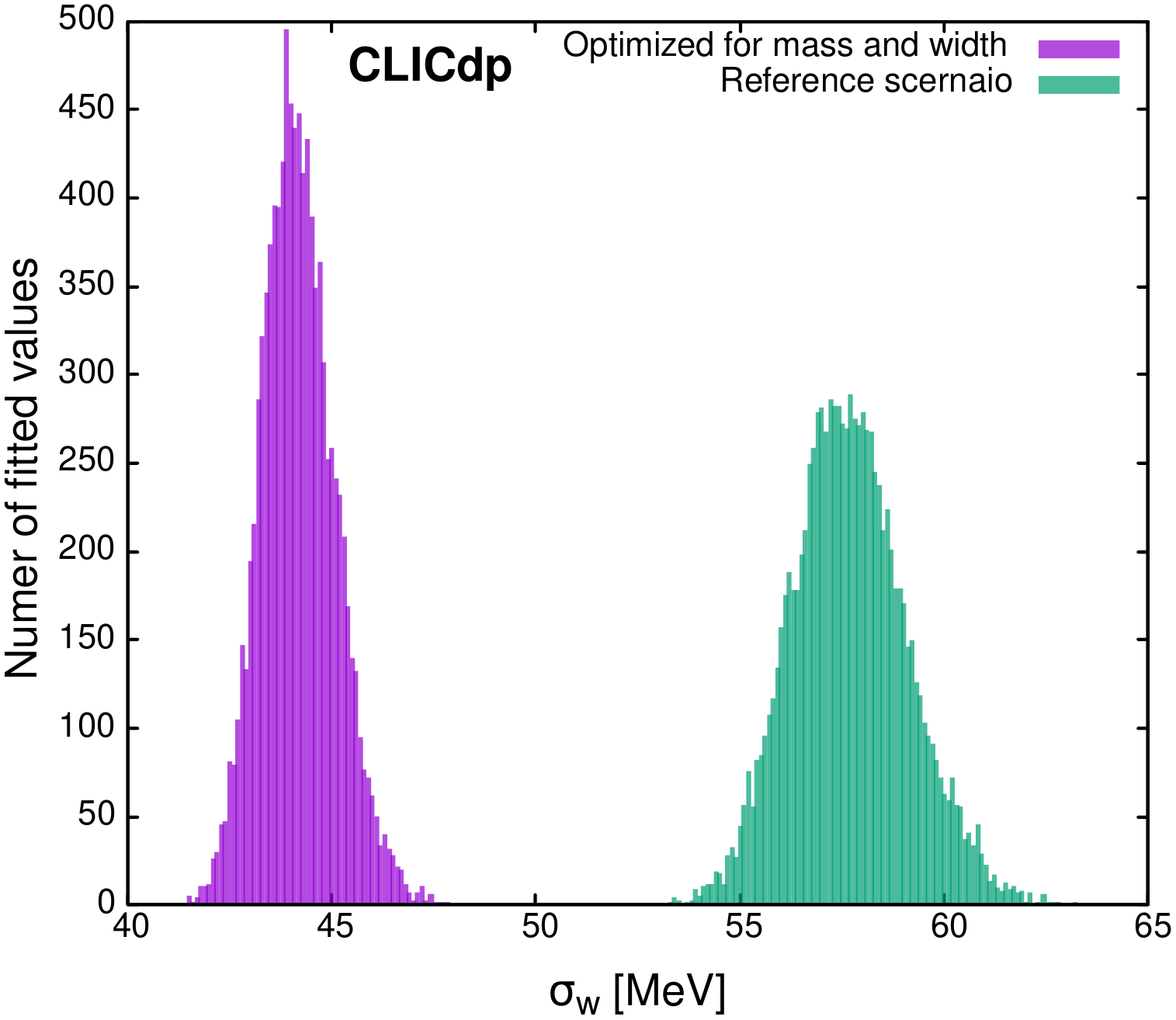}
        \caption{Uncertainty distribution for mass (left) and width (right) measurement, 
        for five-point scan scenario optimised for mass and width measurement 
        (see Fig. \ref{fig:scen}), compared with the distributions for the reference scenario.  }
        \label{fig:histMW}
    \end{figure}
    
Results of the comparison for the five-point scenario are presented in
Fig.~\ref{fig:histMW}.  
Shown are the distributions of uncertainties on the top-quark mass and width.
Results based on large sample of pseudo-experiments confirm
estimates obtained from the optimisation procedure (where each
scenario was evaluated based on three pseudo-experiments only).
For the optimised scenario the average expected mass uncertainty 
is around 26\,MeV, while for width it is around 44\,MeV.
  Moreover,  uncertainty distributions are narrower that those for the
  reference scenario confirming that the fit is very stable and less
  sensitive to the  statistical  fluctuations.

 \begin{figure}[tb]
        \centering
\includegraphics[height=5.7cm]{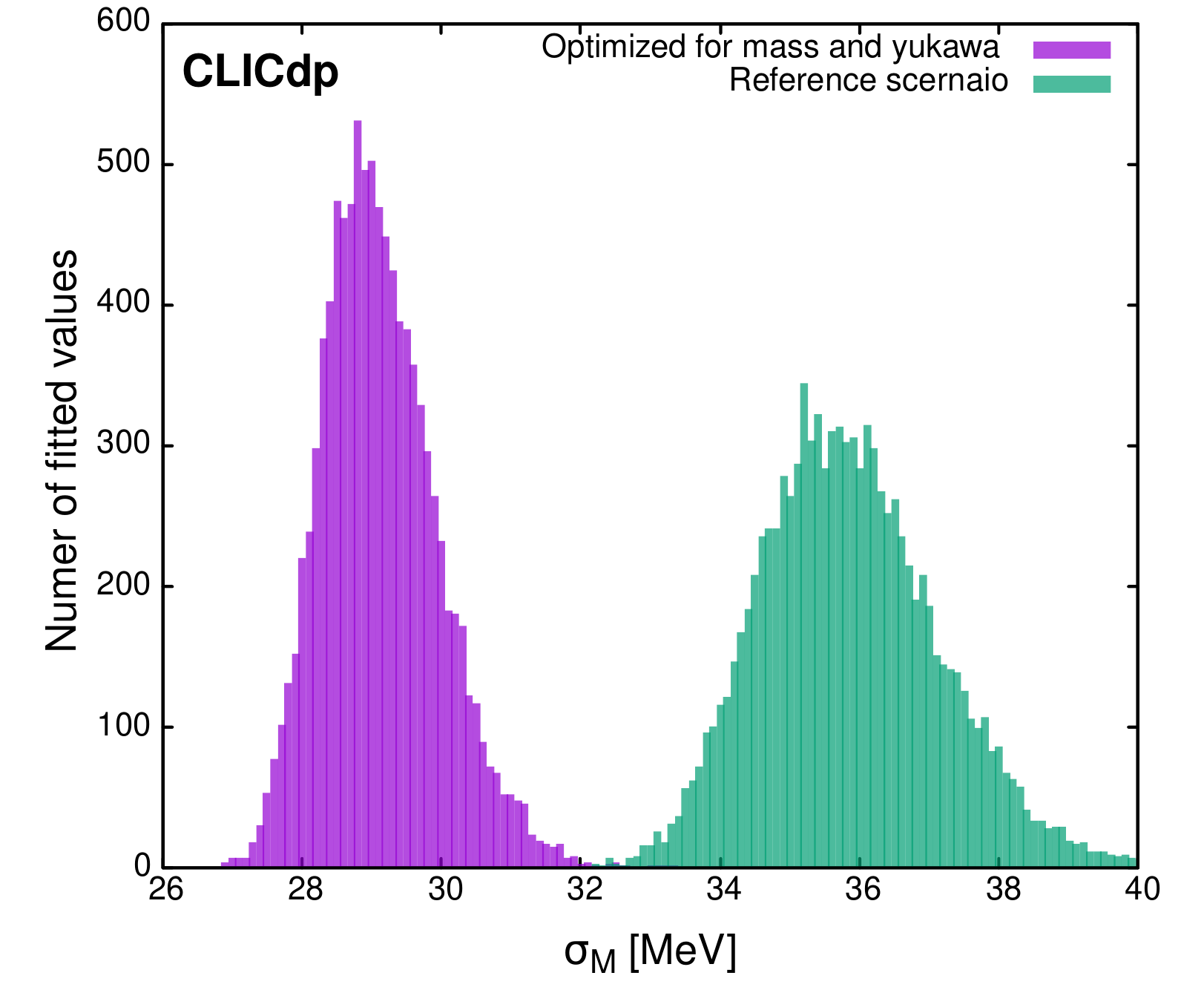}
\hspace{0.5cm}        
\includegraphics[height=5.7cm]{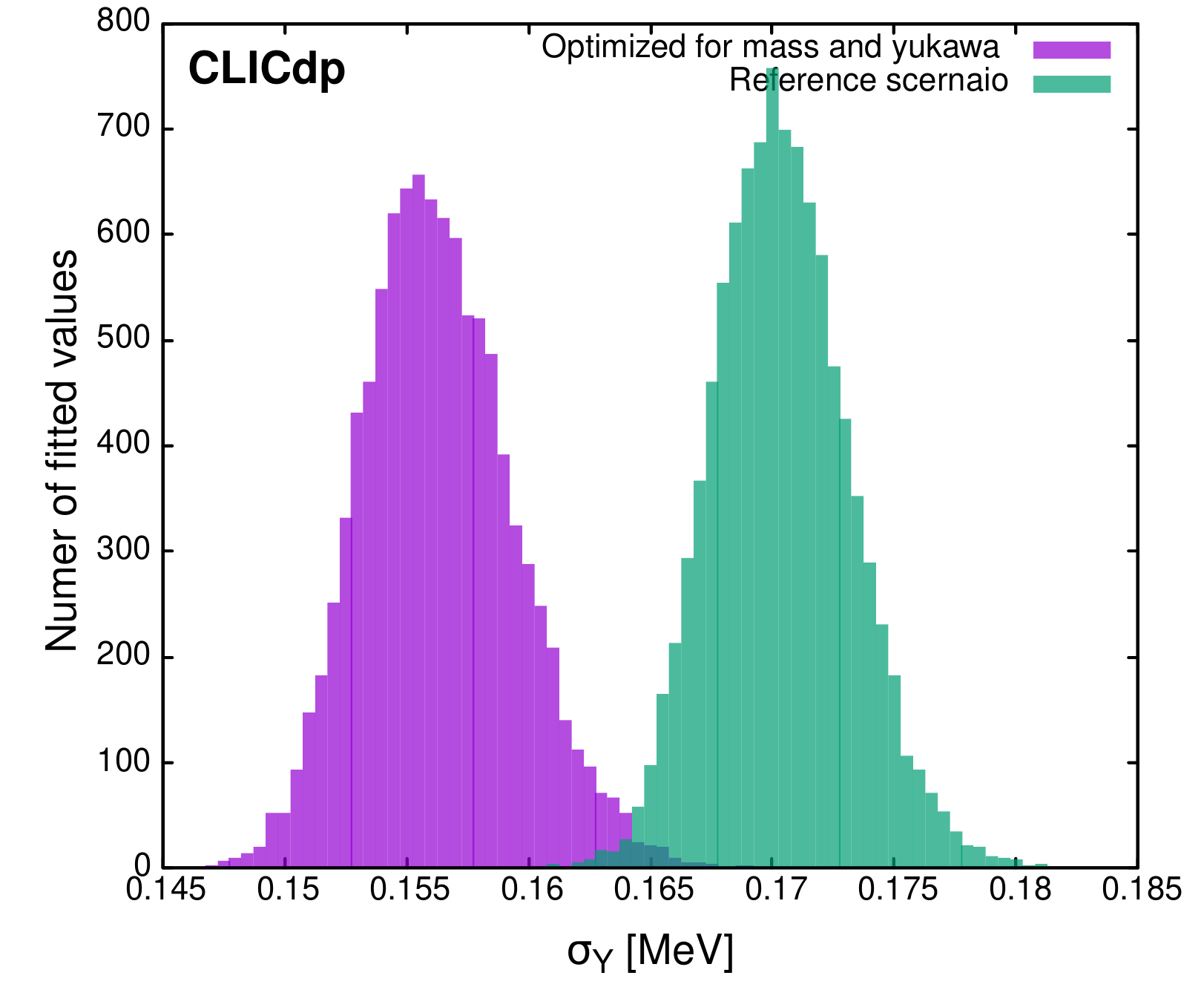}
      \caption{Uncertainty distribution for mass (left) and Yukawa coupling (right) measurement, 
        for 8 point scan scenario optimised for mass and Yukawa coupling measurement 
        (see Fig. \ref{fig:scen}), compared with the distributions for the reference scenario.  }
        \label{fig:histMY}
    \end{figure}
    
For scenario optimised for mass and Yukawa coupling measurement, 
results obtained with large sample of pseudo-experiments shown in
Fig.~\ref{fig:histMY} are again in good agreement with optimisation
results (see Fig.~\ref{fig:multiMY}).
However, uncertainty distribution for the mass
measurement is significantly wider than the one obtained for mass and
width optimised scenario (Fig.~\ref{fig:histMW}).
It is also slightly assymetric, with a larger tail towards high
uncertainty values, but the optimised scenario always gives better
mass measurement precision that the reference one.
Similar tail towards high uncertainty values is also visible for the
Yukawa coupling uncertainty distribution, 
which overlaps slightly with the reference scenario one.
Still, chance of getting measurement from the optimised scan worse 
than in the reference scenario is very small.

\subsection{Impact of the initial mass uncertainty}

 The nominal procedure of pseudo-experiment generation starts from
 selecting the 'true' model parameters, including  the true value of
 top-quark mass.
 However, the mass value will only be known with limited precision
 before the actual experiment takes place, so the position of the scan
 energy points w.r.t. the true mass will not be known.
 To verify how this can affect the expected scan results, fit
 procedure was repeated multiple times for different values of the
 true mass assumed in the generation of the pseudo-experiment data.
 Both the reference scan scenario and the scenario optimised for the
 mass and width measurement were considered with mass variations of up
 to $\pm 0.4$\,GeV from the nominal top-quark mass assumed in the
 optimisation procedure,
 which corresponds to twice the projected future
 experimental uncertainty of the top-quark mass measurement at the HL-LHC
 \cite{Azzi:2019yne}. 
 
\begin{figure}[tb]
     \centering
\includegraphics[height=5.7cm]{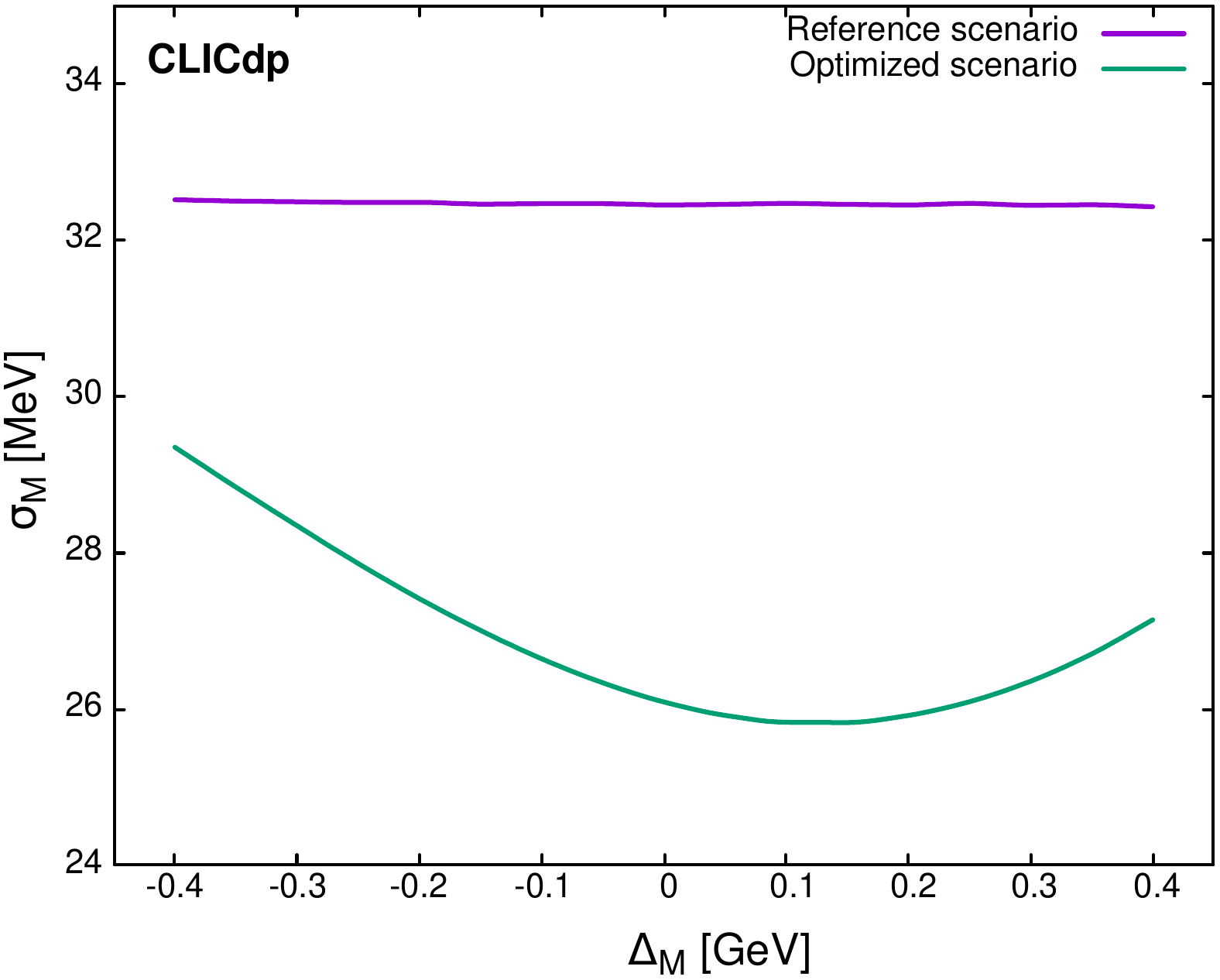}
\includegraphics[height=5.7cm]{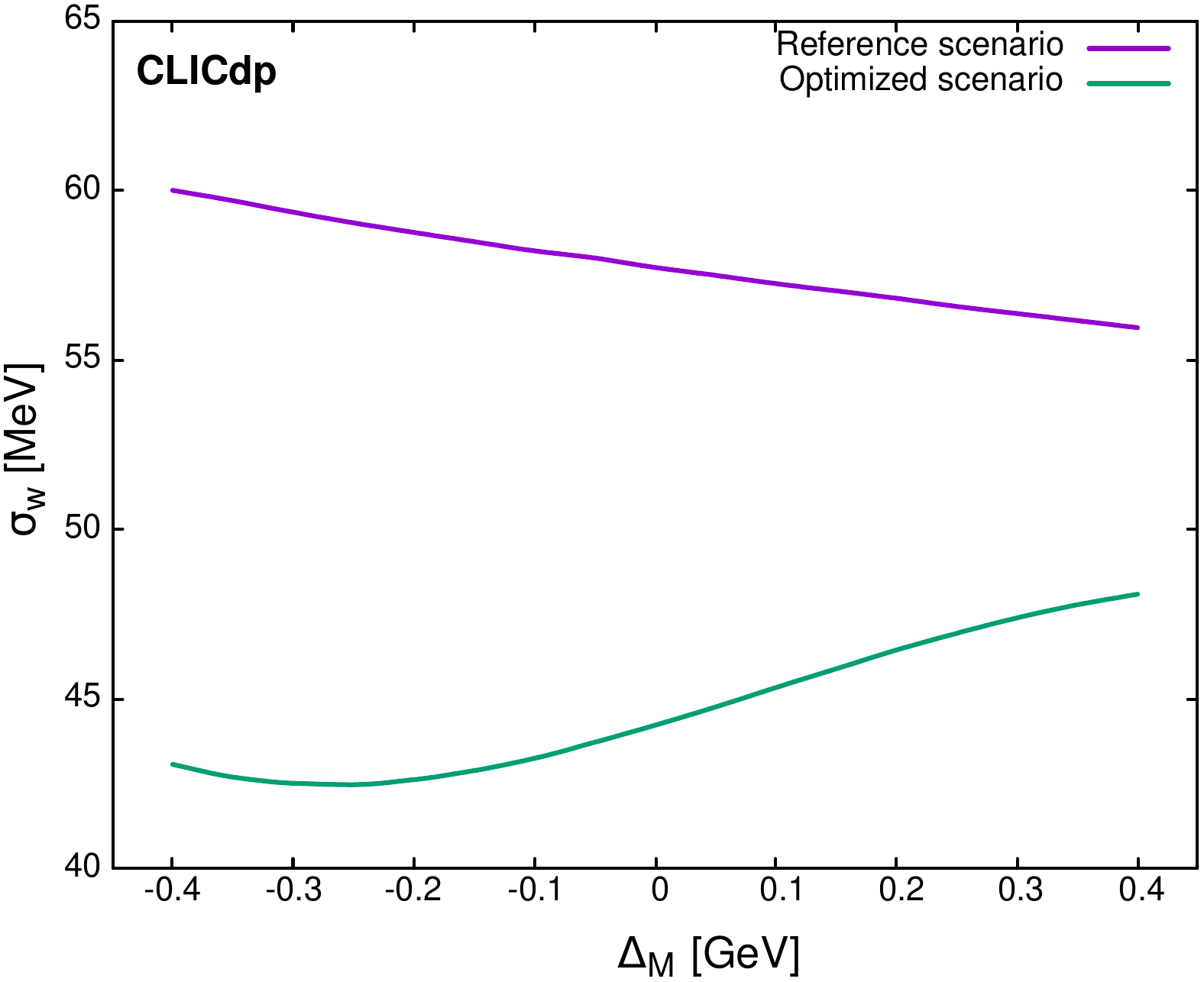}
     \caption{Dependence of the top-quark mass (left) and width
       (right) uncertainties from the 5D fit on the initial top-quark
       mass shift relative to the nominal mass assumed in the
       optimisation procedure. Compared are results obtained for the
       reference scan scenario and the scenario optimised for mass and
       with measurement.
     } 
     \label{fig:Mshift}
 \end{figure}

 The average fitted mass values follow very closely the variation of
 the input mass and no sygnificant deviation is observed for both the
 reference and the optimised scan scenario.
 More pronounced is the impact of the initial mass variation in the
 estimated parameter uncertainties.
 Presented in Fig.~\ref{fig:Mshift} are uncertainties on the top-quark
 mass and width from the 5D fit as a function of the initial top-quark
 mass variation, $\Delta_M$.
 For the reference scenario, the impact of the mass shift on
 the mass uncertainty is negligible, but it does affect the width
 uncertainty. 
 The lowest energy point of the reference  scenario, as defined in
 \cite{clic-top}, is at the  begining of the threshold slope. 
 For negative mass shifts, when the the actual top quark mass is
 lower than the nominal mass, less luminosity is 
 collected in the threshold region sensitive to the width variations
 (see also Fig.~\ref{fig:templates}).
 
 For the optimised scan scenario the best mass measurement precision
 is obtained for small positive mass shift, $\Delta_M \approx 0.1$\,GeV, while
 for negative shifts the uncertainty increases.
 Reversed relation is observed for the width uncertainty, which is
 smallest for negative mass shift of $\Delta_M \approx -0.25$\,GeV and
 increases for the positive shifts.
 This observation is in agreement with results presented in
 Fig.~\ref{fig:single3} indicating that the energy point on the threshold
 slope optimal for the mass measurement is slightly below the point
 optimal for the width measurement.
 Selected in the combined optimisation procedured is the running energy which
 is between the two one-variable choices.
 
 Although both mass and width measurements can be affected by the
 uncertainty in the reference top-quark mass assumed for planning the
 threshold scan, the precision expected from the optimised scan is
 superior to the one of the reference scenario for the 
 initial mass variations considered.
 This suggests that the mass determination precision od 200\,MeV
 expected from the HL-LHC running~\cite{Azzi:2019yne} should already be
 sufficient for initial optimisation of the \ttbar threshold scan at CLIC.

\subsection{Renormalisation scale variation}

In the study \cite{clic-top} the dominant systematic uncertainty in
the top-quark mass measurement from the threshold scan was attributed
to the missing higher orders in theory calculations.
This uncertainty was estimated with the variation of the QCD
renormalisation scale parameter, $\mu$, used as an input to \qqthr
calculations (see Tab.~\ref{tab:qq_par}) and was found to be around
40\,MeV.
Unfortunately, this type of uncertainty can not be included in the presented
analysis framework.
Default value of the renormalisation scale used in \qqthr
calculations corresponds to the maximum value of the top-pair
production cross section at the threshold \cite{Simon:2019axh} and the
linear approximation used in the proposed fit procedure, resulting in
parabolic dependence of the $\chi^2_{\alpha}$ value on the model
parameters (see Eq.~(\ref{eq:para})) is no longer valid.
Therefore, the renormalisation scale variation was not included in the
fit procedure but was considered separately, as described below.
One has to realise that the scale variation is only an effective way
of estimating the theoretical uncertainties and does not correspond to
any parametric uncertainty which could be defined in terms of the probability
density function. 
One can also expect that the theoretical calculations will still
improve before the threshold scan is eventually performed.

\begin{figure}[tb]
     \centering
\includegraphics[height=6cm]{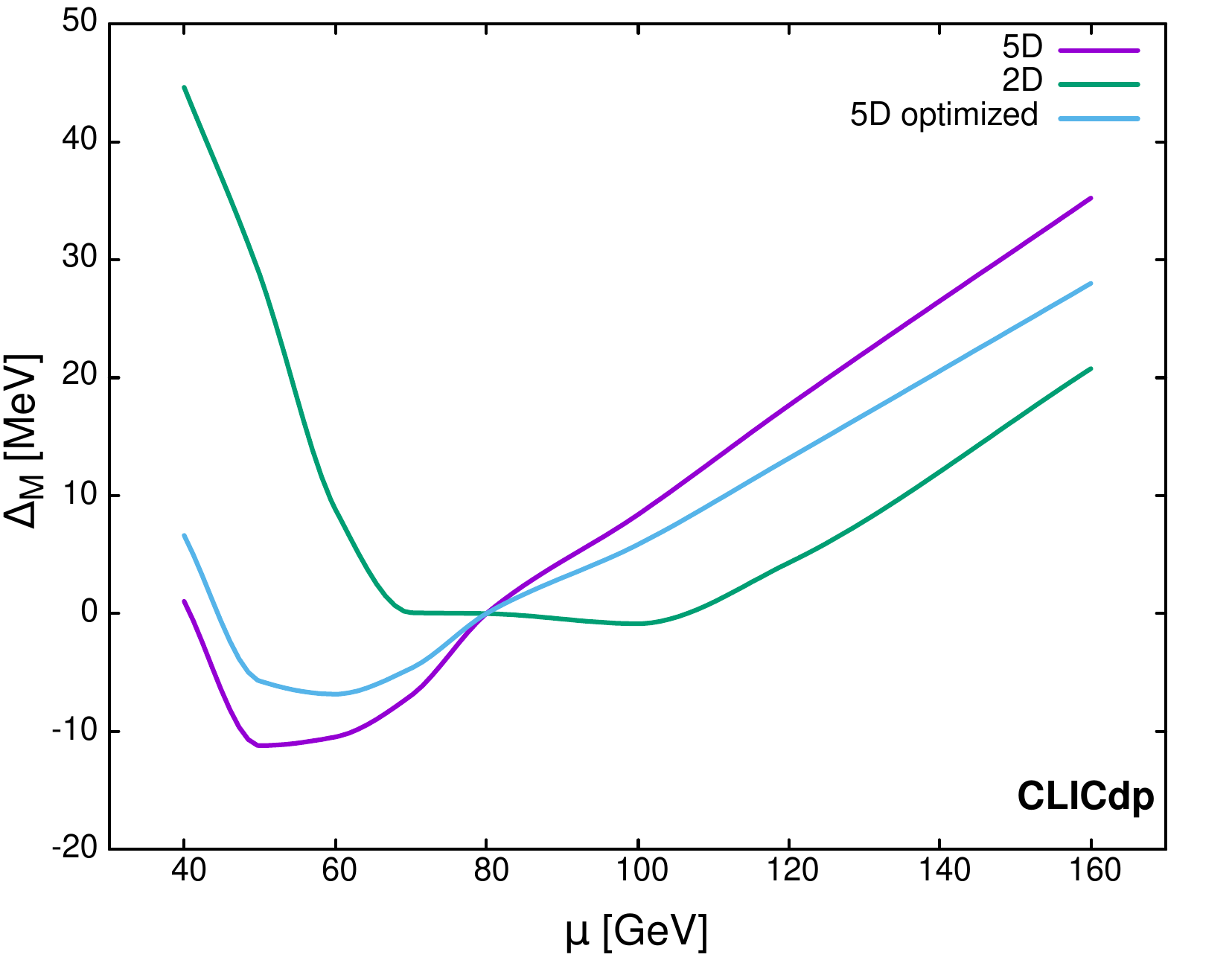}
     \caption{Dependence of the shift in the average top-quark mass
       extracted from the fit to the threshold scan data on the
       variation of the QCD renormalisation scale parameter, $\mu$,
       used for the generation of the pseudo-experiment data sets.
       Compared are results obtained from the 2-D and 5-D fits to the
       reference scan scenario and the 5-D fit to the scan optimised
       for mass and with measurement.
     } 
     \label{fig:mu}
 \end{figure}

To quantify the dependence of the fit results on the assumed QCD
renormalisation scale, fit procedure was repeated multiple times for
different values of $\mu$ assumed in the generation of the
pseudo-data sets while the nominal scale value $\mu = 80$\,GeV was
always used in the fit procedure.
Dependence of the extracted top-quark mass on the input renormalisation
scale, for three different fit configurations, is presented in
Fig.~\ref{fig:mu}. 
For the two parameter fit to the reference scan scenario data, maximum
mass shift of about 40\,MeV is observed, consistent with results of
\cite{clic-top}.
The dependence of the mass shift on the renormalisation scale is
modified significantly when 5-D fit to the reference scan scenario is
considered and the resulting uncertainty estimate (corresponding to the
scale variation by factor of 2) is reduced to about 35\,MeV.
Sensitivity to the renormalisation scale is further reduced, by about
20\%,  for the optimised scan scenario.
Expected uncertainty from the theory calculations can be estimated to
about 28\,MeV.

\subsection{Influence of the luminosity spectra}

All results presented so far were obtained assuming 100\,fb$^{-1}$ of
data collected during the top threshold scan at CLIC.
This luminosity is not sufficient to reach statistical uncertainty
below 20\,MeV, which was considered as a goal in the previous study
\cite{clic-top}.
However, fit uncertainties can be significantly reduced, if
more luminosity is collected at the threshold. 
Expected measurement precision is also sensitive to the assumed
shape of the luminosity spectra.

\begin{figure}[tb]
        \centering
        \includegraphics[height=7cm]{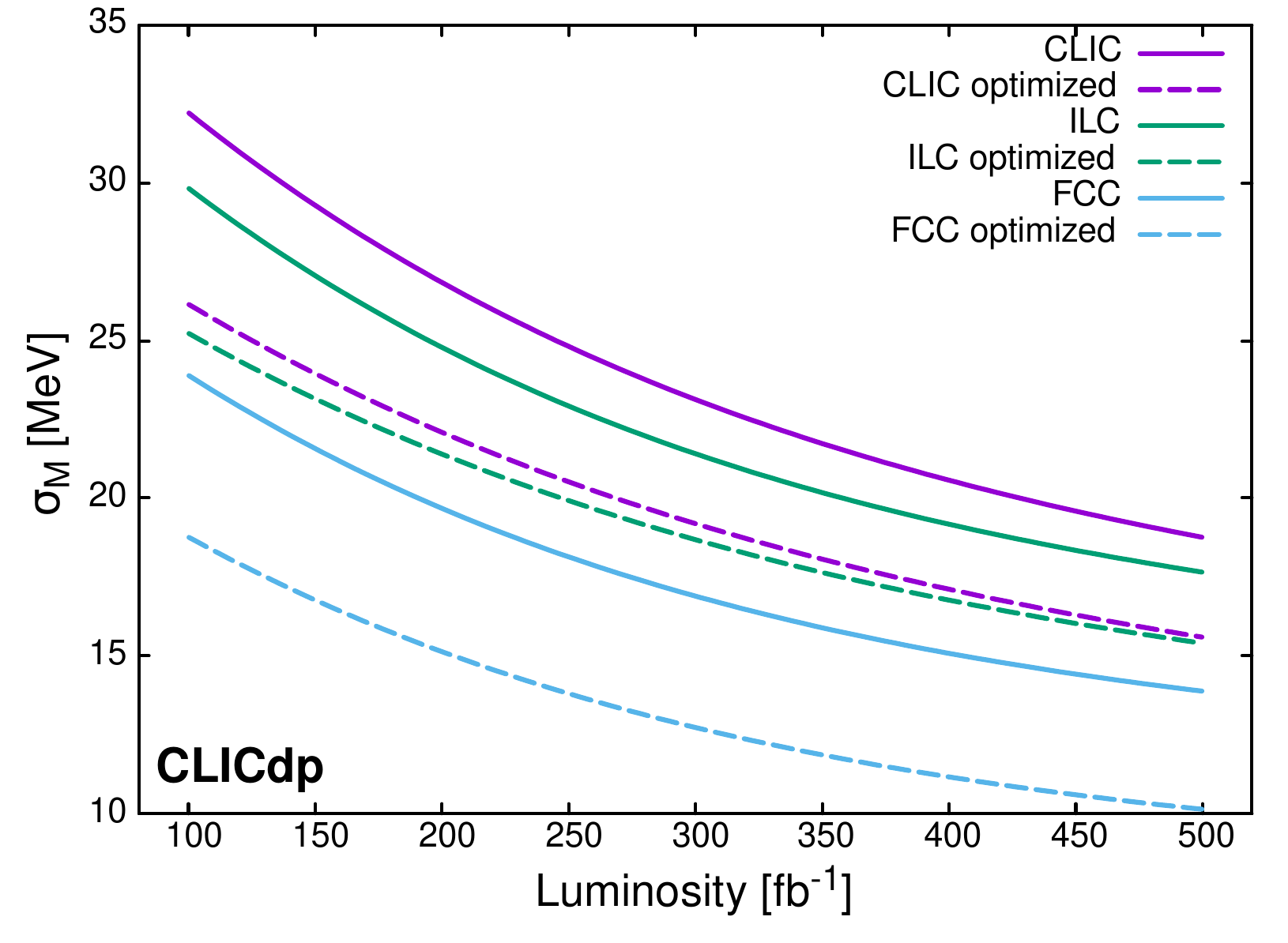}
        \includegraphics[height=7cm]{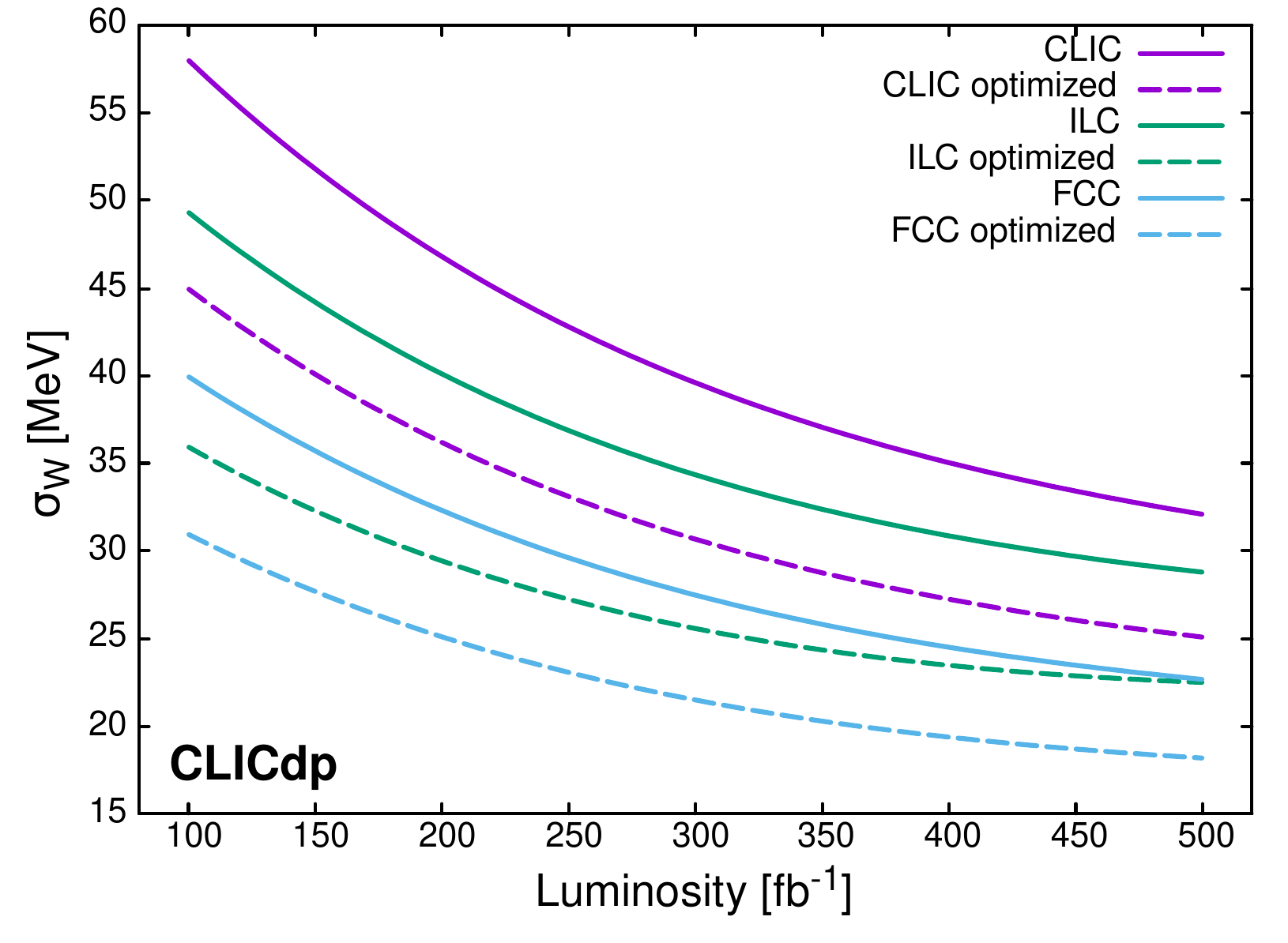}
\caption{Expected uncertainty on the top-quark mass (top) and width
  (botom) from the baseline scan scenario (solid lines) and from
  scenarios optimised for mass and  width measurement (dashed lines), as a
  function of the total scan luminosity.
  CLIC results (magenta) are compared with results of the same analysis and
  optimisation procedure but assuming ILC (green) or FCCee (cyan)
  luminosity spectra. 
} 
        \label{fig:my_label}
\end{figure}
Dependencies of the expected mass and width uncertainties on the
integrated luminosity of the threshold scan are shown in
Fig.~\ref{fig:my_label}.
CLIC estimates are compared with results of the same analysis and
optimisation procedure applied when assuming ILC~\cite{ilc} or
FCC-ee~\cite{fccee} luminosity spectra.
To make comparison even more realistic, 'true' top quark mass is
varied in the pseudo-data generation procedure to model its initial
uncertainty of 200\,MeV \cite{Azzi:2019yne}. 
Note that the choice of the optimal scenario does change with the change
of luminosity spectra.
For ILC spectra, about 90\% of scenarios from the last generation of
mass and width optimisation include 6 energy points.
Six point optimised scenario was therefore selected for the comparison.
Similarly, seven-point scenario was used for \mbox{FCC-ee}, as this was
the final number of energy points in all scenarios of the last
generation.
Note that only the luminosity spectra was changed for this comparison;
reconstruction efficiency and background level estimated from the CLIC
study~\cite{Seidel}  were used for all spectra and the possible
influence of the beam polarisation was not considered.

Due to additional constraints imposed on the data normalisation,
background contribution, the strong coupling constant and Yukawa
coupling, the expected uncertainties decrease slower than with
$\sqrt{\cal L}$.   
In order to achieve 20\,MeV mass uncertainty, about 400\,fb$^{-1}$ of
data is required  with reference running scenario at CLIC and about 
200\,fb$^{-1}$ at FCC-ee.
The optimised scenarios can reach this level of precision already with
about 250\,fb$^{-1}$ at CLIC and below 100\,fb$^{-1}$ at FCC-ee.
This shows clear advantage of optimised scenarios, as they can provide
same precision as the reference scenario with the luminosity lower by
roughly a factor of 2.  

As described above, results presented in Fig.~\ref{fig:my_label} were
obtained for the most general 5D fit approach.
When the ``SM constrained'' 3D fit is performed to the same scan data,
with top-quark width and Yukawa coupling taken from the SM
predictions, mass uncertainty from the scan scenario optimised for
CLIC is reduced by about 2\,MeV for the whole range of the integrated
luminosity values.  
It is also interesting to notice that very similar mass uncertainties
are obtained after the optimisation procedure for CLIC and ILC luminosity 
spectra.
On the other hand, much higher width precision is obtained for the ILC.
This is due to the fact that, compared to the five point scenario
optimal for CLIC, the scenario optimised for ILC includes additional scan
point in the region below the threshold (at around 341\,GeV) improving
the width determination, but reducing the luminosity collected at the higher
energy points (more relevant for mass determination). 

Presented results confirm that the optimisation procedure based on
Genetic algorithm can be used to propose different running scenarios.
It is possible to improve precision of top-quark mass
determination by about 20\%, also improving the measurement of the
top-quark width or top Yukawa coupling at the same time.
With optimised running scenario, the statistical uncertainties on 
mass and width are similar to the uncertainties expected for the
reference scenario with doubled integrated luminosity (200\,fb$^{-1}$).


    \section{Conclusions}
    
Complementary fit and optimisation procedures have been developed 
for the top-quark threshold scan analysis at CLIC.
The new fit procedure is more flexible that the one used in the
previous study~\cite{clic-top} and allows to include all relevant
model parameters as well as additional constraints on model
parameters, coming eg. from earlier measurement, and constraints on
data normalisation.
For the baseline scan scenario assumed at CLIC, with 100\,fb$^{-1}$ of
integrated data luminosity, top quark mass can be measured with
uncertainty of 32\,MeV assuming the current uncertainty of the strong
coupling constant, relative uncertainty of the Yukawa coupling of 0.1
and the background normalisation to be better than 2\%.
At the same time, cross section contribution from the top Yukawa
coupling can be confirmed with statistical significance higher that
$5\sigma$. 
To improve the mass determination precision to below 25\,MeV the
strong coupling constant would need to be known with uncertainty below
0.0003.   
 
Optimisation procedure based on non dominated sorting genetic
algorithm II has been applied to the top-quark pair-production threshold scan.
Each measurement scenario (set of energy points with total equally
distributed luminosity of 100 fb\,$^{-1}$) is considered a genotype
and results of the fit procedure constitute a phenotype.
Implementation of the genetic evolution, starting from the baseline
scenario, includes random mutations: mixing of parent genotypes,
possibility to drop or gain a new (random) chromosome (scan
point). Stable optimisation results are obtained for population size
of 2000 and number of generation of 30.
Using single and multi objective optimisation, it was shown that
optimisation procedure can reduce the mass uncertainty by up to 20\%.
With the proposed procedure, reduction of the mass uncertainty from
the top threshold scan fit should be possible, corresponding to 
about a factor of 2 increase in the integrated luminosity.
For the optimised running scenario statistical uncertainty on the
top-quark mass of 20\,MeV can be reached at CLIC with 250\,fb$^{-1}$. 
Scan optimisation results also in the reduced sensitivity to 
      the renormalisation scale variations.

\subsection*{Acknowledgements}

The work was carried out in the framework of the CLIC detector and
physics (CLICdp) collaboration.
We acknowledge partial use of the earlier results from
K.\,Debski~\cite{kacper_debski}.  
We thank collaboration members for fruitful discussions, valuable
comments and suggestions.
The work was partially supported by the National Science Centre
(Poland) under OPUS research projects no. 2017/25/B/ST2/00496
(2018-2021).

\bibliographystyle{JHEP}

\bibliography{top_jhep}

\end{document}